\documentclass[a4paper,10pt]{article}
\usepackage[utf8]{inputenc}

\date{}
\usepackage{natbib}

\usepackage[]{graphicx}

\usepackage[]{color}

\definecolor{fgcolor}{rgb}{0.345, 0.345, 0.345}
\newcommand{\hlnum}[1]{\textcolor[rgb]{0.686,0.059,0.569}{#1}}%
\newcommand{\hlopt}[1]{\textcolor[rgb]{0,0,0}{#1}}%
\newcommand{\hlstd}[1]{\textcolor[rgb]{0.345,0.345,0.345}{#1}}%
\newcommand{\hlkwb}[1]{\textcolor[rgb]{0.69,0.353,0.396}{#1}}%
\newcommand{\hlkwc}[1]{\textcolor[rgb]{0.333,0.667,0.333}{#1}}%
\newcommand{\hlkwd}[1]{\textcolor[rgb]{0.737,0.353,0.396}{\textbf{#1}}}%

\usepackage{framed}
\makeatletter
\newenvironment{kframe}{%
 \def\at@end@of@kframe{}%
 \ifinner\ifhmode%
  \def\at@end@of@kframe{\end{minipage}}%
  \begin{minipage}{\columnwidth}%
 \fi\fi%
 \def\FrameCommand##1{\hskip\@totalleftmargin \hskip-\fboxsep
 \colorbox{shadecolor}{##1}\hskip-\fboxsep
     \hskip-\linewidth \hskip-\@totalleftmargin \hskip\columnwidth}%
 \MakeFramed {\advance\hsize-\width
   \@totalleftmargin\z@ \linewidth\hsize
   \@setminipage}}%
 {\par\unskip\endMakeFramed%
 \at@end@of@kframe}
\makeatother

\definecolor{shadecolor}{rgb}{.97, .97, .97}
\definecolor{messagecolor}{rgb}{0, 0, 0}
\definecolor{warningcolor}{rgb}{1, 0, 1}
\definecolor{errorcolor}{rgb}{1, 0, 0}
\newenvironment{knitrout}{}{} 

\usepackage{alltt}

\usepackage{moreverb}
\usepackage{url}
\usepackage{grffile}
\usepackage{lipsum}
\usepackage[UKenglish]{isodate}
\usepackage{booktabs} 
\usepackage{tablefootnote}

\usepackage{hyperref}
\hypersetup{
	colorlinks=true,
	urlcolor=blue,
	citecolor=blue,
	linktoc=all,
	linkcolor=blue}

\newcommand\BibTeX{{\rmfamily B\kern-.05em \textsc{i\kern-.025em b}\kern-.08em
T\kern-.1667em\lower.7ex\hbox{E}\kern-.125emX}}
\newcommand\MiKTeX{{\rmfamily M\kern-.05em \textsc{i\kern-.025em K}\kern-.08em
T\kern-.1667em\lower.7ex\hbox{E}\kern-.125emX}}
\newcommand\PracTeX{{\rmfamily P\kern-.05em \textsc{r\kern-.025em a\kern-.025em
c}\kern-.08em
T\kern-.1667em\lower.7ex\hbox{E}\kern-.125emX}}

\usepackage{listings}
\usepackage{xspace}
\usepackage{amsmath}
\usepackage{amssymb, mathtools}
\usepackage{mathrsfs}
\usepackage{graphicx}
\graphicspath{{figures/}}
\usepackage{caption}
\usepackage{cleveref}
\usepackage{multirow}
\usepackage{xcolor}
\colorlet{mygreen}{green!60!gray}
\colorlet{myred}{red!60!gray}
\def\yess{\textcolor{mygreen}{\textsf{yes}}}
\def\nos{\textcolor{myred}{\textsf{no}}}

\usepackage[margin=1in]{geometry}

\def\d{\mathrm{d}}
\def\simind{\stackrel{\mathrm{ind}}{\sim}}
\def\simiid{\stackrel{\mathrm{iid}}{\sim}}

\newcommand{\X}{\mathbb{X}}

\newcommand{\edr}{\mathrm{e}}
\newcommand{\ddr}{\mathrm{d}}

\DeclareMathOperator*{\argmin}{arg\,min}
\def\VI{\mathcal{VI}\xspace}

\def\BNPdensity{\pkg{BNPdensity}\xspace}
\def\R{\textbf{R}\xspace}

\crefname{equation}{}{} 

\usepackage{nth}
\usepackage{enumerate} 
\usepackage{enumitem}
\usepackage{cleveref}
\newcommand{\code}[1]{\texttt{#1}}
\newcommand{\pkg}[1]{\textsf{#1}} 

\usepackage{etoolbox} 
\makeatletter
\preto{\@verbatim}{\topsep=-10pt \partopsep=-10pt }
\makeatother

\usepackage[acronym]{glossaries}
\makeglossaries
\newacronym{ssd}{SSD}{Species Sensitivity Distribution}
\newacronym{cec}{CEC}{Critical Effect Concentration}
\newacronym{ecx}{$\mathrm{EC}_x$}{Effect Concentration at $x\%$}
\newacronym{lc50}{$\mathrm{LC}_{50}$}{Lethal Concentration $50\%$}
\newacronym{ec50}{$\mathrm{EC}_{50}$}{Effect Concentration $50\%$}
\newacronym{hc5}{$\mathrm{HC}_{5}$}{Hazardous Concentration for $5\%$ of the Species}
\newacronym{iid}{\emph{iid}}{identically and independently distributed}
\newacronym{cpo}{CPO}{conditional predictive ordinate}
\newacronym{loo}{LOO}{Leave-One-Out}
\newacronym{bnp}{BNP}{Bayesian Nonparametrics}
\newacronym{rivm}{RIVM}{National Institute for Public Health and the Environment}
\newacronym{kde}{KDE}{Kernel Density Estimate}
\newacronym{CDF}{CDF}{Cumulative Distribution Function}
\newacronym{mcmc}{MCMC}{Markov chain Monte Carlo}
\newacronym{MH}{M-H}{Metropolis--Hastings}
\newacronym{NRMI}{NRMI}{Normalised Random Measures with Independent Increments}
\newacronym{DPM}{DPM}{Dirichlet Process Mixture}
\newacronym{NGG}{NGG}{Normalised Generalised Gamma}

\title{\BNPdensity:  Bayesian nonparametric  mixture modeling in \R}

\author{
Julyan ARBEL$^1$, Guillaume KON KAM KING$^2$\thanks{Corresponding author}, Antonio LIJOI$^3$,\\ 	Luis Enrique NIETO-BARAJAS$^4$, Igor PR\"UNSTER$^3$ \\\\
\and
$^1$Université Grenoble Alpes \\ Inria, CNRS
 LJK, Grenoble INP \\ 38000 Grenoble, France\\
 \texttt{julyan.arbel@inria.fr}
\and 
$^2$Université Paris-Saclay \\INRAE, MaIAGE, \\ 78350, Jouy-en-Josas, France \\
\texttt{guillaume.kon-kam-king@inrae.fr}
\and
$^3$Department of Decision Sciences\\
BIDSA, Bocconi University, Milan, Italy\\\texttt{igor.pruenster@unibocconi.it} \\
\texttt{antonio.lijoi@unibocconi.it} 
    \and
$^4$Department of Statistics\\
ITAM, Mexico\\
\texttt{lnieto@itam.mx} 
}

\begin{document}

\maketitle

\begin{abstract}
Robust statistical data modelling under potential model mis-specification often requires leaving the parametric world for the nonparametric.
In the latter, parameters are infinite dimensional objects such as functions, probability distributions or infinite vectors.
In the Bayesian nonparametric approach, prior distributions are designed for these parameters, which provide a handle to manage the complexity of nonparametric models in practice.
However, most modern Bayesian nonparametric models seem often out of reach to practitioners, as inference algorithms need careful design to deal with the infinite number of parameters.
The aim of this work is to facilitate the journey by providing computational tools for Bayesian nonparametric inference.
The article describes a set of functions available in the \R package \BNPdensity in order to carry out density estimation with an infinite mixture model, including all types of censored data.
The package provides access to a large class of such models based on normalized random measures, which represent a generalization of the popular Dirichlet process mixture.
One striking advantage of this generalization is that it offers much more robust priors on the number of clusters than the Dirichlet.
Another crucial advantage is the complete flexibility in specifying the prior for the scale and location parameters of the clusters, because conjugacy is not required.
Inference is performed using a theoretically grounded approximate sampling methodology known as the Ferguson \& Klass algorithm.
The package also offers several goodness of fit diagnostics such as QQ-plots, including a cross-validation criterion, the conditional predictive ordinate.
The proposed methodology is illustrated on a classical ecological risk assessment method called the Species Sensitivity Distribution (SSD) problem, showcasing the benefits of the Bayesian nonparametric framework.
\end{abstract}

\section{Introduction}
\label{sec:intro}

\R \citep{rcoreteam} is often cited by Bayesian statisticians as their favorite programming language due to the many packages that provide tools for Bayesian inference. 
The general program for Bayesian inference \pkg{BUGS} \citep{gilks1994language} has been available for a couple of decades, with interfaces in \R. 
Since then, additional software has been developed to make that language more accessible to the users, for instance \pkg{OpenBUGS} \citep{thomas2006making}, \pkg{JAGS} \citep{plummer2003jags}, and \pkg{Stan} \citep{stan-software:2015}.
All three can be accessed directly from \R by respectively using  \pkg{R2OpenBUGS}/\pkg{R2WinBUGS} \citep{sturtz2005r2winbugs},  \pkg{rjags}  \citep{rjags2019plummer}, \pkg{runjags}  \citep{denwood2016runjags}, and  \pkg{rstan}  \citep{rstan-software}.  
Programs for specific fields of Bayesian statistics have appeared in recent years, for instance 
\pkg{bspmma} \citep{burr2010bspmma} for meta-analysis using \gls{DPM} models, 
\pkg{DPpackage} \citep{jara2007applied,jara2011DPpackage}, a bundle of functions for Bayesian nonparametric models,  
\pkg{BNPmix} \citep{canale2019bnpmix}, a set of functions for density estimation with Dirichlet process and Pitman--Yor mixing measures via marginal algorithms,
\pkg{PReMiuM} \citep{Liverani2013} for profile regression using the Dirichlet process, 
\pkg{Biips} \citep{Biips} for Bayesian inference via particle filtering, 
\pkg{Bayesian Regression} \citep{karabatsos2015menu} for Bayesian nonparametric regression. 
Packages \pkg{mcclust} \citep{Scrucca2016}, \pkg{mcclust.ext}  \citep{wade2018bayesian} and \pkg{GreedyEPL} \citep{Rastelli2018} provide point estimation and credible sets for Bayesian cluster analysis. 
The interested reader may refer to the CRAN Task View on Bayesian Inference 
for an extensive list of \R packages dedicated to Bayesian statistics (see Section~\ref{sec:pkg-comparison} for a more detailed discussion of \R packages for Bayesian density estimation).

Robust statistical data modeling under potential model mis-specification often requires relaxing parametric assumptions for nonparametric assumptions.
In \gls{bnp}, parameters are infinite dimensional objects such as functions, probability distributions or infinite vectors.
Prior distributions are designed for these parameters, which provide a handle to manage the complexity of nonparametric models in practice.
However, the applicability of \gls{bnp} models, for data analysis, depends on the availability of user-friendly software. This is because \gls{bnp} models typically require complex representations, which may not be immediately accessible to non-experts. 
This work focuses on inference of densities with mixture models \citep{handbook2018}. 
The purpose of the present paper is to introduce and describe an extensive revamping of the \BNPdensity package, originally presented in \citet{barrios2013modeling}. 
The package is programmed in \R, and is available from the Comprehensive \R Archive Network (CRAN) at \url{https://CRAN.R-project.org/package=BNPdensity}.
To the best of our knowledge, \BNPdensity is the first \R package which implements \gls{bnp} density models including all types of censored data (left-, right- and interval-censored data), under a general  specification of \gls{bnp} priors called normalised generalised gamma processes \citep{lijoi2007controlling,barrios2013modeling}.
The improvements to the package cover various aspects.
Notably, careful profiling and re-writing of some critical parts of the code, along with the
use of the \R bytecode compiler, yielded a 4-fold decrease of the running time of the algorithm.
Drawing on the flexibility of the algorithm to use non-conjugate prior, we also implemented a range of popular new priors on the scale parameter of the clusters such as the half-Cauchy \citep{Gelman2006,Chung2015}, the truncated Gaussian and the uniform distributions.
We also revised the truncation method in the algorithm, intended to deal with the infinite dimensional random measures in the \gls{bnp} model, to include recent contributions by \cite{arbel2017moment}.
These provide a better and principled control of the truncation approximation.
Moreover, we extended \BNPdensity to include all types of censored data (right-, left- or interval-censored data).
To leverage on the clustering properties of \gls{bnp} mixture models, we interfaced \BNPdensity with other packages to estimate the optimal clustering from posterior samples and provided cluster visualisation tools.
We also implemented functions to compute prior distributions on the number of mixture components, for various processes, to better inform prior specification.
Finally, we added several new functions for graphical model checking, assessing \gls{mcmc} convergence and parallel computation.

The paper is organised as follows. 
We start with a concise overview of Bayesian nonparametric mixture models for density estimation in Section~\ref{sec:scope}, along with our strategy for posterior inference and a description of the recent improvements to \BNPdensity. 
We then describe the package and its general syntax in Section~\ref{sec:package}, including some simple examples, and provide in Section~\ref{sec:pkg-comparison} a comprehensive comparison of the features and functionalities offered in three \R packages dedicated to \gls{bnp} density estimation, namely: \BNPdensity, \pkg{BNPmix}, and \pkg{DPpackage}. 
We then conclude with a case study in Section~\ref{sec:case_study}.

\section{Bayesian nonparametric density estimation} \label{sec:scope}

This section aims at providing a concise
review of the statistical model used in the \BNPdensity package. 
As the name suggests, the focus of the package is density estimation based on \gls{bnp} priors, including all types of censored data. 
The density model used is a mixture model \citep{handbook2018}, where the mixing measure is a \gls{bnp} prior, thus leading to an infinite mixture model. 

The most widely used \gls{bnp} mixture model for density estimation is the \glsfirst{DPM} model due to \citet{lo1984class}.
 Generalisations of the \gls{DPM} correspond to allowing the mixing distribution to be any discrete nonparametric prior. 
 A large class of such prior distributions is obtained by normalising  increasing additive processes \citep{sato1999levy}.
  The normalisation step, under suitable conditions, gives rise to so-called \gls{NRMI} as introduced in \citet{rlp2003}. See also
 \citet{barrios2013modeling}.
 
We focus on a class of NRMIs that are obtained by normalising the increments of a generalised gamma process  \citep{brix1999generalized} proposed in \citet{lijoi07}, which enjoy analytical tractability and include
many well-known priors as special cases. 
Generalised gamma processes are discrete random measures $\tilde \rho$ of the form
\begin{equation}\label{eq:GG}
	\tilde \rho = \sum_{i=1}^\infty J_i\delta_{\boldsymbol{\theta}_i},
\end{equation}
where the weights $J_i$ do not sum to one and are such that $\sum_{i\ge1}J_i<\infty$ almost surely, while the location parameters $\boldsymbol{\theta}_i$ are sampled iid from a  measure $P_0$, a probability distribution on the parameter space $\Theta$. 
In what follows, $P_0$ is considered as diffuse.
$(J_i, \boldsymbol{\theta}_i)$ are the points of a Poisson process with mean intensity: 
\begin{equation}
\label{eq:ngg} \nu(\d v,\d \boldsymbol{\theta}) =\frac{\edr^{-\kappa v}}{\Gamma(1-\gamma)  v^{1+\gamma}}\,\d v\, \alpha
P_0(\d \boldsymbol{\theta}),
\end{equation}
which depends on parameters $\kappa\geq0$ and $\gamma\in[0,1)$ such that $(\kappa,\gamma)\neq (0,0)$. 
The measure $\nu$ in~\eqref{eq:ngg} characterises $\tilde{\rho}$ and is often referred to as the Lévy intensity. 
The base measure is $\alpha P_0$, where $\alpha>0$. 
The corresponding
generalised gamma \gls{NRMI}, obtained by normalising the generalised gamma process as $\tilde P(\,\cdot\,) := \tilde \rho(\,\cdot\,)/\tilde \rho(\Theta)$ will be denoted as $\tilde P \sim\operatorname
{NGG}(\alpha,\kappa,\gamma; P_0)$. 
This class of priors contains as special cases the Dirichlet process which is a
$\operatorname
{NGG}(\alpha, 1, 0; P_0)$ process, the normalised inverse Gaussian
(N-IG) process \citep{lijoi05}, which corresponds to a $\operatorname{NGG}(1,
\kappa,1/2; P_0)$ process, and the \mbox{N-stable} process \citep{kingman1975random}
which arises as $\operatorname{NGG}(1, 0 ,\gamma; P_0)$.

We now describe the mixture model in more detail. 
We consider a density kernel $k(  \cdot   \mid  \boldsymbol{\theta})$ mixed with respect to $\tilde P \sim\operatorname
{NGG}(\alpha,\kappa,\gamma; P_0)$ 
thus obtaining the random mixture density 
\begin{equation}\label{eq:random_density}
	\tilde f(x)=\int_{\Theta}
k(x\mid\boldsymbol{\theta}) \tilde P(\ddr\boldsymbol{\theta}).
\end{equation}
This can equivalently be written in
a hierarchical form as
\begin{equation}\label{eq:mixt}
\begin{split}
	X_i \mid \boldsymbol{\theta}_i & \simind k( \cdot \mid
\boldsymbol{\theta}_i),\quad i=1,\ldots,n,
\\
\boldsymbol{\theta}_i\mid\tilde P & \simiid \tilde P,\quad i=1,\ldots,n,
\\
\tilde P & \sim \operatorname
{NGG}(\alpha,\kappa,\gamma; P_0).
\end{split}
\end{equation}
Details on possible choices for the kernel $k$ and the base measure $P_0$ are provided in Section~\ref{sec:package}, while in Section~\ref{sec:pkg-comparison} we argue that conjugacy is not required in this setting.

We denote by $f_0$ the density with respect to the Lebesgue measure of the NGG base measure $P_0$ on $\Theta$. 
When $P_0$ depends on a
further hyperparameter $\boldsymbol{\phi}$, we use the notation $f_0(  \cdot  \mid 
\boldsymbol{\phi})$. Using the \code{MixNRMI2} function corresponds to the specification of a
nonparametric model for the 
location and scale parameters of the mixture where the mixture parameter $\boldsymbol{\theta}$ takes the form of the vector $(\mu
,\sigma)$. 
In order to distinguish the hyperparameters for
location and scale, we will use the notation $f_0(\mu,\sigma \mid \boldsymbol{\phi}
)=f_{0}^1(\mu \mid \sigma, \varphi) f_{0}^2(\sigma \mid  \varsigma)$. 
In
applications a priori independence between $\mu$ and $\sigma$ is
commonly assumed, and this is indeed a natural assumption for the illustration in \Cref{sec:case_study}.

The most popular uses of mixtures with discrete random probability
measures, such as the one displayed in \eqref{eq:mixt}, relate to
density estimation and data clustering. 
The former can be addressed by
evaluating the posterior expectation of the random density $\tilde f$ defined in~\eqref{eq:random_density}, given a sample $\mathbf{X}=(X_1,\ldots,X_n)^\top$,
\begin{equation}
\hat{f}_n(x)=\textrm{E} \bigl(\tilde f(x)  \mid  \mathbf{X} \bigr) \label{eq:density_est}
\end{equation}
for any $x$.
As for the latter, if $R_n$ is the number of distinct latent values
$\boldsymbol{\theta}_1^*,\ldots,\boldsymbol{\theta}_{R_n}^*$ out of a sample of size~$n$, one can
deduce a partition of the observations such that any two $X_i$ and
$X_j$ belong to the same cluster if the corresponding latent variables
$\boldsymbol{\theta}_i$ and $\boldsymbol{\theta}_j$ coincide. 
Then, it is interesting to
determine an estimate $\hat{R}_n$ of the number of clusters into which
the data are grouped, along with the clustering structure. 
For details on clustering estimation in our setting, see Section~\ref{sec:clustering}.

In the next subsection, we show how to solve all estimation problems
with a posterior sampling algorithm.

\subsection{Posterior sampling via a conditional Gibbs sampler}\label{sec:posterior-sampling}

According
to the terminology of \cite{papaspiliopoulos2008retrospective}, posterior sampling methods for \gls{bnp} mixture models can be divided into two classes: marginal and conditional methods. 
Marginal methods, such
as \cite{escobar1995bayesian, MacEachern1998, neal2000markov}, integrate out
the the infinite-dimensional component \eqref{eq:GG} of the hierarchical
model and sample from the marginal distribution of the remaining variables.
Conditional methods work directly on \eqref{eq:mixt} and must solve the problem of sampling the trajectories of
an infinite-dimensional random element. 
However, they
allow inference on the latent random measure $\tilde P$, for instance on the jump sizes. 
An example of conditional method, which nicely fits our framework, is the Ferguson and Klass algorithm. 
Unlike 
marginal samplers, it allows for estimating non-linear functionals of 
the underlying posterior distribution, such as credible intervals.
Here we sketch the conditional algorithm implemented in \BNPdensity which
allows to draw posterior simulations from mixtures based on a general \gls{NRMI} (a very thorough description of the algorithm can be found in \citealp{barrios2013modeling}).
It works equally well regardless of whether the kernel $k$ and $P_0$
form a conjugate pair  and readily yields
credible intervals. 
The algorithm is an implementation of the posterior characterisation of \glspl{NRMI} provided in \cite{james2009posterior}.

For $n$ observations $\mathbf{X}=(X_1,\ldots,X_n)^\top$ in  $\mathbb{R}$,  we consider the random distribution function induced by $\tilde \rho$, 
$$\tilde{M} :=
\left\{
\tilde{M}(\boldsymbol{s})=
(
\tilde{\rho}((-\infty, s_{1}]), 
\cdots,
\tilde\rho((-\infty, s_{n}]
)^\top,
\quad \boldsymbol{s}=(s_{1}, \ldots, s_{n})^\top \in \mathbb{R}^{n} 
\right\}
.
$$
For the implementation of the Gibbs sampling scheme,
we use the distributions of $[\tilde{M}  \mid  \mathbf{X}, \boldsymbol{\theta}]$ and $[\boldsymbol{\theta}  \mid  \mathbf{X}, \tilde{M}]$. 
Due to conditional independence properties, the
conditional distribution of $\tilde M$, given $\mathbf{X}$ and $\theta$, does not depend on $\mathbf{X}$, that is, $[\tilde{M}  \mid  \mathbf{X}, \boldsymbol{\theta}]=[\tilde{M}  \mid  \boldsymbol{\theta}]$. 
Thanks to Theorem 1 in \cite{barrios2013modeling} \citep[originating in][]{james2009posterior}, the posterior distribution function $[\tilde{M}  \mid  \boldsymbol{\theta}]$ can be  characterised as a mixture in terms of a latent variable $U$, that is through the distributions $[\tilde M  \mid  U, \boldsymbol{\theta}]$ and $[U  \mid  \boldsymbol{\theta}]$. 
Thus, the Gibbs sampler uses the following conditional distributions:
\begin{enumerate}
\item $[U  \mid  \boldsymbol{\theta}]$: sampling the latent variable $U$ conditionally on the latent parameters $\boldsymbol \theta$, where $U$ follows the distribution:
\begin{equation}\label{eq:U-post}
f_{U  \mid  \mathbf{X}}(u) \propto u^{n-1}(u+\kappa)^{r \gamma-n} \exp \left\{-\frac{a}{\gamma}(u+\kappa)^{\gamma}\right\}.
\end{equation}
Sampling $U$ is performed via a \gls{MH} step with a gamma proposal distribution $\operatorname{ga}\left(\delta, \delta / u^{[t]}\right)$ centered at the previous $U$ value $u^{[t]}$ with a tuning parameter $\delta$ controlling the coefficient of variation. An adaptive version of the  \gls{MH} algorithm \citep{roberts2009examples} without the tuning parameter is also implemented in the package, and proposed with the option \code{adaptive=TRUE}. It uses a log-transformation of the random variable $U$. Note that the target density~\eqref{eq:U-post} not being log-concave, ergodicity cannot be proven as in \cite{roberts2009examples}. Nevertheless, the adaptive version appears to offer superior performance in practice.
 \item $[\tilde M  \mid  U, \boldsymbol{\theta}]$: simulating the infinite dimensional process conditionally on the parameters and the latent variable $U$. 
 This is performed using the \cite{ferguson1972representation} algorithm.
 According to to Theorem 1 in \cite{barrios2013modeling}, the conditional distribution of $\tilde M$ is  composed of two parts, a part without fixed points of discontinuity $\tilde M^*$ which can be expressed as an infinite sum of random jumps occurring at random locations and a part with fixed points of discontinuity, or in other words: $\tilde{M}(\boldsymbol{s}) =\tilde{M}^{*}(\boldsymbol{s})+\sum_{j=1}^{R_n} J_{j}^{*} \mathbb{I}_{(-\infty,\boldsymbol{s}]}(\boldsymbol{\theta}_{j}^{*})$ where the $\boldsymbol{\theta}_{j}^{*}$, $j=1,\ldots,R_n$ denote the $R_n$ distinct parameters among $\boldsymbol{\theta}_1,\ldots,\boldsymbol{\theta}_n$ and where $(-\infty,\boldsymbol{s}]=\{\boldsymbol{x}\in\mathbb{R}^n:x_i\leq s_i,i=1,\ldots,n\}$. 
 In the infinite sum:
 \begin{equation} \label{eq:infinite_series}
 \tilde{M}^{*}(\boldsymbol{s}) = \sum_{j=1}^{\infty} J_{j} \mathbb{I}_{(-\infty,\boldsymbol{s}]}(\boldsymbol{\vartheta}_{j}),
\end{equation}
 the $J_j$s are obtained by inverting the relation $\xi_j=N(J_j)$, where $\xi_1, \xi_2, \ldots$ are jump times of  a standard Poisson process of unit rate, that is $\xi_{1}, \xi_{2}-\xi_{1}, \ldots \stackrel{\mathrm{iid}}{\sim} \operatorname{ga}(1,1)$, with
 \begin{equation}
 	N(v)=\frac{a}{\Gamma(1-\gamma)} \int_{v}^{\infty} \mathrm{e}^{-(\kappa+u) x} x^{-(1+\gamma)} \mathrm{d} x,
 \end{equation}
while the jumps $\boldsymbol{\vartheta}_{j} = (\boldsymbol{\vartheta}_{j}^{(1)},\ldots,\boldsymbol{\vartheta}_{j}^{(n)})^\top$ are sampled from the base measure $P_0$.
The jumps $J_j^*$ at the
fixed locations $\boldsymbol{\theta}_j^*$ are gamma distributed:
\begin{equation}
 f_{j}^{*}(v)=\frac{(\kappa+u)^{n_{j}-\gamma}}{\Gamma\left(n_{j}-\gamma\right)} v^{n_{j}-\gamma-1} \mathrm{e}^{-(\kappa+u) v},
\end{equation}
where $n_j$ are the multiplicities, i.e. the number of $\boldsymbol{\theta}_j$ equal to $\boldsymbol{\theta}_j^*$.
A fundamental merit of Ferguson
and Klass' representation, compared to similar algorithms, is the fact that the random heights $J_i$ are obtained in a descending order.
Therefore, one can
truncate the series in \eqref{eq:infinite_series} at a certain finite index $Q$ to be decided via a moment-matching criterion (see \Cref{sec:moment_matching}).
This also guarantees that the
highest jumps are not left out.
 \item $[\boldsymbol{\theta} \mid  \mathbf{X}, \tilde M]$: resampling the latent cluster parameters given the data and the random measure. The support of
the conditional distribution of $\boldsymbol{\theta}_i$ are the set of locations
$\{\bar{\boldsymbol{\vartheta}}_{j}\}_{j=1}^\infty = \{\boldsymbol{\theta}_1^*,\ldots,\boldsymbol{\theta}_{R_n}^*,\boldsymbol{\vartheta}_1,\ldots\}$ with associated jump $\{\bar{J}_{j}\}_{j=1}^\infty = \{J_1^*,\ldots,J_{R_n}^*,J_1,\ldots\}$ of $\tilde M$,
 \begin{equation}
  f_{\boldsymbol{\theta}_{i}  \mid  X_{i}, \tilde{M}}(\boldsymbol{s}) \propto \sum_{j} k\left(X_{i}  \mid  \boldsymbol{s}\right) \bar{J}_{j} \delta_{\bar{\boldsymbol{\vartheta}}_{j}}(\mathrm{d} \boldsymbol{s}).
 \end{equation}
Simulating from this conditional distribution when an approximation with a finite number of jumps has been determined is
straightforward: one just needs to evaluate the right-hand side of the expression above and normalise.

\item Updating the hyperparameters of $P_0$. We only put a prior on the hyperparameters for the location parameters, and found this to have a higher impact.
Assuming a priori independence between location and scale parameters of the clusters, the conditional posterior distribution on the hyperparameters given the data and the rest of the parameters only depends on the distinct location parameters.
A simple way to proceed is thus to consider a prior conjugate to the base measure.
\end{enumerate}

We also include a resampling of the unique values of the cluster parameters via a \gls{MH} step to avoid the `sticky clusters effect', as suggested in \cite{bush}.

We devote the next section to explaining the moment-matching criterion used for truncation in the second conditional, which is a recent addition to the package \BNPdensity.

\subsection{Moment-matching criterion}\label{sec:moment_matching}

\gls{NGG} priors are infinite dimensional objects that are obtained by normalising a generalised gamma process. 
Concrete implementation of \gls{NGG} priors 
requires to truncate the random series \eqref{eq:GG} at some level denoted $Q$, which results in some truncation error. 
Previous implementation of the package used to appeal to a relative error index, that we will denote $e_Q = \sum_{i>Q}J_i\delta_{\boldsymbol{\theta}_i}$, based on the jumps themselves. 
We improve on this approach, by implementing the methodology proposed by \cite{arbel2017moment} which relies on a moment-based evaluation of the error, denoted by $\ell_M$. 
One of the main contributions of  \cite{arbel2017moment} is to warn that relying on the  relative error index $e_M$ can lead to overly optimistic conclusions in terms of approximation, especially for large values of the discount parameter $\gamma$. 

To be more specific, consider $K$ moments of the total mass of the CRM $\tilde \rho(\X)=\sum_{i=1}^\infty J_i$, denoted by $\boldsymbol{m}_K = (m_1,\ldots,m_K)^\top$. 
Such moments have a simple expression in terms of the cumulants, which are themselves available in  closed form, see for instance Table 1 in \cite{arbel2017moment}. Thus, these exact moments can be computed and compared with their empirical counterparts obtained with the Ferguson \& Klass algorithm \citep{ferguson1972representation}. 

In order to make this methodology applicable, one needs to propose the truncation level $Q(\ell)$ required to achieve a given approximation $\ell$. 
Such map $Q(\ell)$  only depends on the \gls{NGG} parameters and can be computed once-for-all and distributed with the package. For reference, see the moment matching error $\ell(Q)$ and the map  $Q(\ell)$ respectively displayed in Figures 1 and 2 of \cite{arbel2017moment}. Ferguson and Klass posterior sampling based on such a prescribed number of jumps $Q(\ell)$ is computationally more efficient than having to iteratively compute the relative error $e_Q$ as done in the previous package version.

\subsection{Clustering estimation}\label{sec:clustering}

We focus here on the problem of estimating a data clustering from the Bayesian posterior inference conducted so far. 
This is a long standing problem in Bayesian statistics \citep[see for instance][]{dahl2006model,lau2007bayesian}. 
Enumerating all partitions is practically not feasible, which typically requires resorting to approximations.

Many ad-hoc procedures have been devised in the literature. However, as noted by \citet{dahl2006model}, it seems counter-intuitive to apply an ad-hoc clustering method on top of a model which itself produces clusterings. 
We adopt instead a fully Bayesian route by undertaking clustering on decision-theoretic grounds. 
We consider a loss function $L$ and propose a Bayesian point estimator $\hat c$ for a clustering obtained as an argument which minimises the posterior expected loss given data $\mathbf{X}$ 
\begin{align}\label{eq:clustering_point_estimator}
\hat c = \argmin_{c^\prime}\sum_{c}L(c^\prime,c)\pi(c\mid \mathbf{X}),
\end{align}
where $\pi(c\mid \mathbf{X})$ is the posterior distribution of clustering $c$. 
Often considered in the literature, the posterior mode is an example of such a Bayesian estimator, based on the very crude 0-1 loss function.
When $n$ is large, an MCMC sample from the posterior generally hardly visits twice the same clustering, thus rendering the empirical mode of the MCMC output very sensitive to the initialisation of the chain and of very limited validity in practice. 
Manifestly, many other loss functions can be considered and expected to perform better than the 0-1 loss. 
One particular choice of a loss function  stands out from these in best estimating the number of groups in a clustering. 
It is called the variation of information, denoted by $\VI$, which is a loss function firmly established in information theory \citep{meila2007comparing,wade2018bayesian}. 
The variation of information between two clusterings is defined as the sum of their information (their Shannon entropies) minus twice the information they share. 
Simulations indicate that the variation of information is a sensible choice: when other losses such as the Binder loss \citep{binder1978bayesian} typically tend to overestimate the number of clusters, the variation of information instead seems to consistently recover it (see for instance the simulated examples, and more specifically Figures 6 to 8, of \citealp{wade2018bayesian}). 

An asset of the approach presented in \citet{wade2018bayesian} is that it rests on a greedy search algorithm to determine the minimum loss clustering of~\eqref{eq:clustering_point_estimator}. 
Starting from the MCMC output, this greedy approach explores the space of partitions and is not restricted to those visited by the MCMC chain to find the optimum. 
We include the possibility to  estimate the optimal clustering using both the $\VI$ loss and Binder's loss, along with other loss functions, within \BNPdensity by adding an optional dependence to \pkg{GreedyEPL}.
Note that clustering estimation is also available for censored data, although graphical representation is more tricky (see also the legend to \Cref{fig:clustering_Carbaryl_cens}).

\begin{minipage}{\textwidth}
\begin{knitrout}\scriptsize
\definecolor{shadecolor}{rgb}{0.969, 0.969, 0.969}\color{fgcolor}\begin{kframe}
\begin{alltt}
\hlkwd{data}\hlstd{(acidity)}
\hlstd{out} \hlkwb{<-} \hlkwd{MixNRMI2}\hlstd{(acidity)}
\hlstd{clustering} \hlkwb{=} \hlkwd{compute_optimal_clustering}\hlstd{(out)}
\hlkwd{plot_clustering_and_CDF}\hlstd{(out, clustering)}
\end{alltt}
\end{kframe}
\end{knitrout}
\end{minipage} 
\begin{center}
 \includegraphics[width=0.5\textwidth]{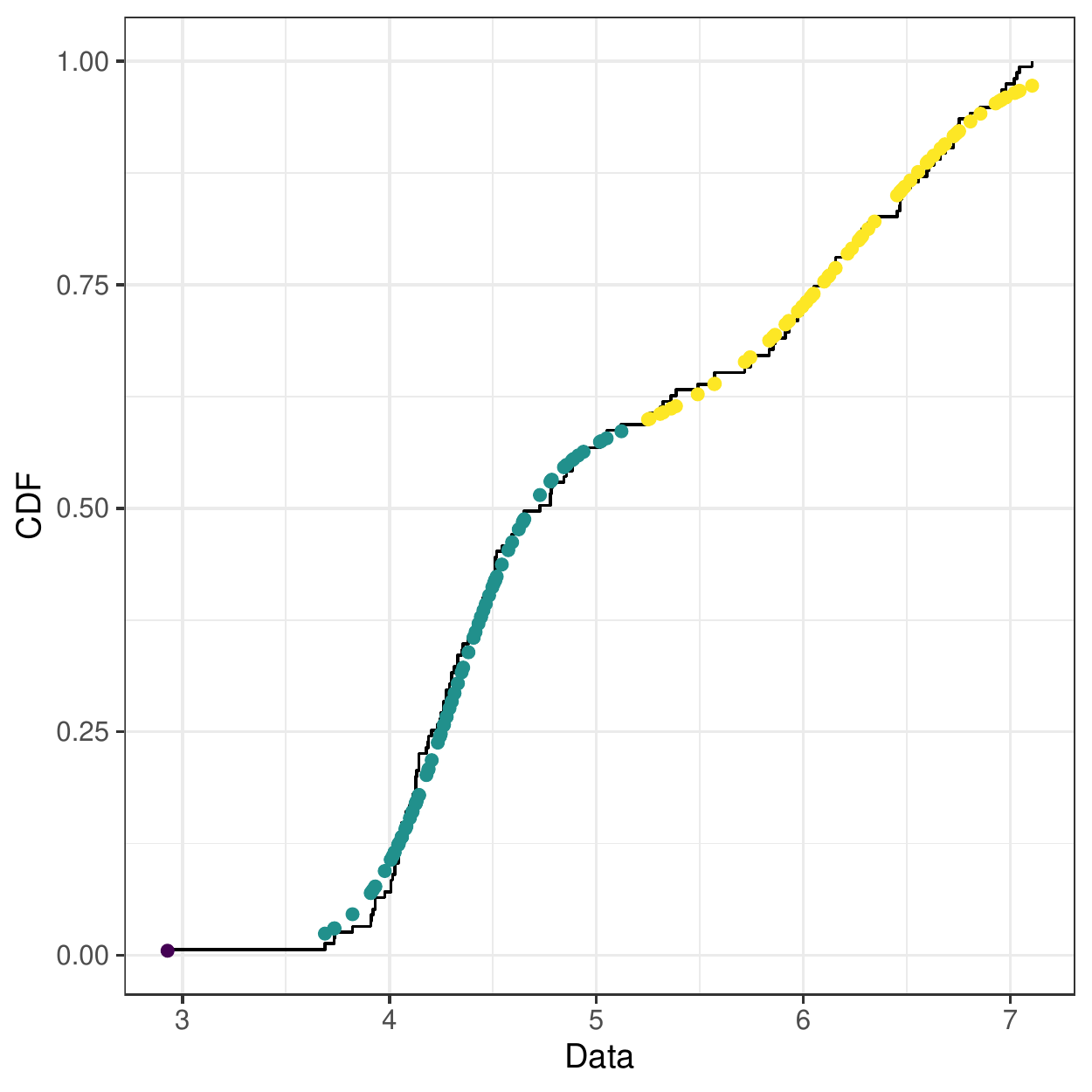}
\captionof{figure}{Visualisation of the clustering induced by the \gls{bnp} mixture model, for the \code{acidity} dataset. 
The solid line represents the empirical \gls{CDF}, dots represent data points. 
The abscissa of each point is its value, the ordinate is the value of the estimated \gls{CDF} at that point.
Each colour denotes the cluster estimated by minimising the $\VI$ loss function.
\label{fig:clustering_example}}
\end{center}

\section{Package description}\label{sec:package}

The implementation of \pkg{BNPdensity} package is available from the Comprehensive \R Archive Network (CRAN) at \url{https://CRAN.R-project.org/package=BNPdensity}. 
Fitting a model with \pkg{BNPdensity} starts with calling one of the two functions, \texttt{\hlstd{MixNRMI1}} or \texttt{\hlstd{MixNRMI2}}, or their versions for censored data. 
The function \texttt{\hlstd{MixNRMI1}} fits a semiparametric mixture model where all components have a common scale parameter $\sigma$ with  an independent parametric prior, $\sigma \sim P_\sigma$, while 
\texttt{\hlstd{MixNRMI2}} is devoted to fully nonparametric mixtures of \textit{location and scale parameters}:
\begin{equation*}
\begin{split}
	X_i \mid \theta_i, \sigma_i & \simind k( \cdot \mid
\theta_i, \sigma_i),\quad i=1,\ldots,n,
\\
(\theta_i, \sigma_i)\mid\tilde P & \simiid \tilde P,\quad i=1,\ldots,n,
\\
\tilde P & \sim \operatorname
{NGG}(\alpha,\kappa,\gamma; P_0).
\end{split}
\end{equation*}

\begin{figure}
\centering
\includegraphics[width = 0.5\textwidth]{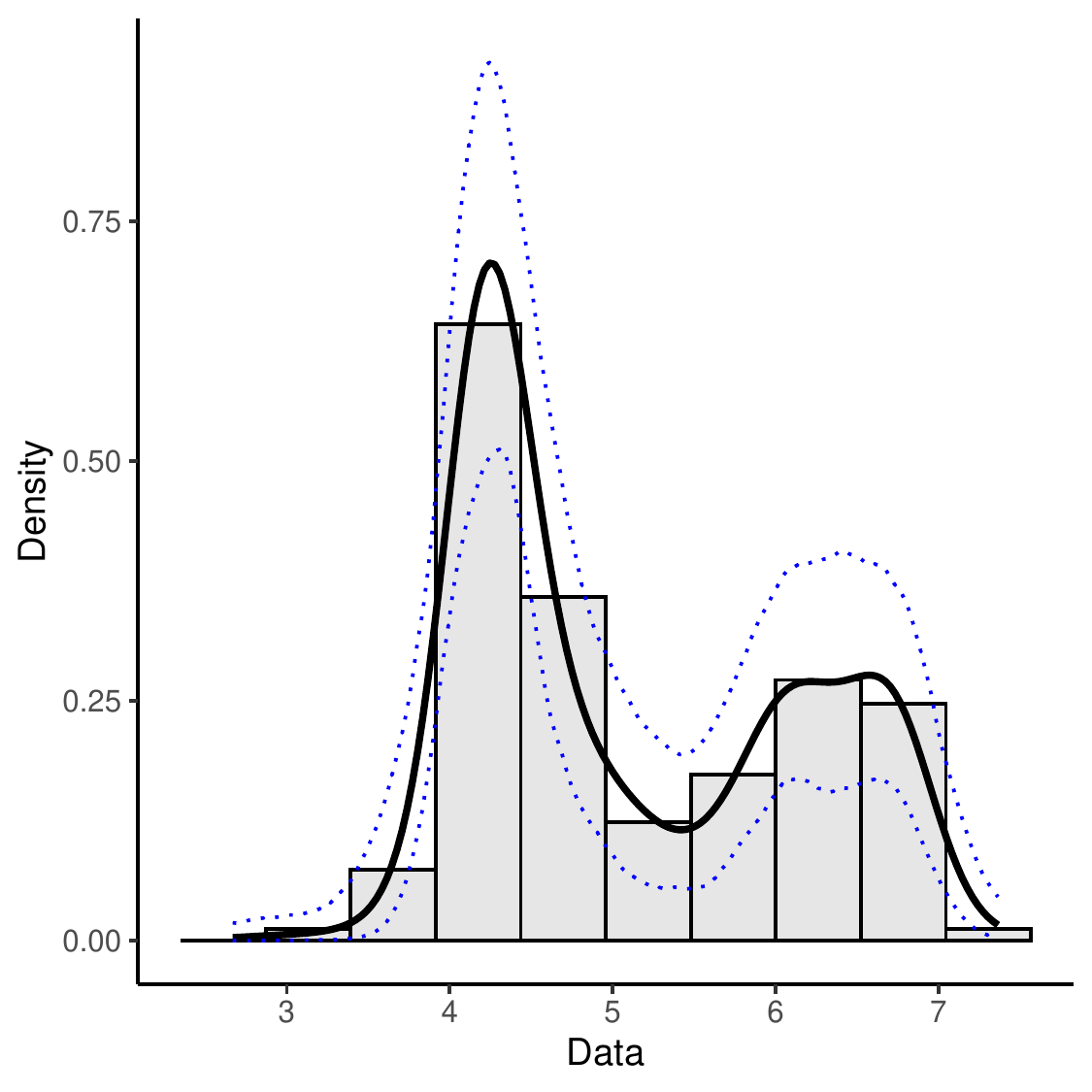}
\caption{Density estimate (solid black line), 95\% credible interval (blue dotted line) and histogram of the \texttt{\hlstd{acidity}} data fitted with a semiparametric model. 
Figure obtained using the command \texttt{\hlkwd{plot}\hlstd{(out)}}.}
\label{fig:density-plot}
\end{figure}

Data and prior parameters are passed to the model function as arguments. 
The \texttt{\hlstd{MixNRMIx}} functions also take a number of arguments to choose the \gls{bnp} model, the mixture kernels, a variety of priors and tuning parameters for the Markov chain Monte Carlo sampling algorithm. 
The main arguments of the model functions are presented below.
\begin{itemize}[noitemsep,topsep=0.5pt]
\item \texttt{\hlstd{distr.k}}: Integer number identifying the \textbf{mixture kernel} $k$. 
Five kernels parameterised by their location and scale are implemented: a Gaussian or double exponential kernel for real data, a gamma or lognormal kernel for positive data and a beta kernel for data on the unit interval.
The flexibility of this choice is afforded by the specific algorithm used in \pkg{BNPdensity}.
\item \texttt{\hlstd{distr.py0}}: Integer number identifying the base measure $P_0$ on the location parameters. Three choices are available, which are constrained by the conjugate prior we place on the hyperparameters of $P_0$: Gaussian, gamma and beta. 
Additional arguments can be used to tune the shape of the base measure.
\item \texttt{\hlstd{distr.py0}}, \texttt{\hlstd{distr.pz0}}: Integer number identifying the base measure $P_0$ on scale parameters. 
For the semiparametric model (\texttt{\hlstd{MixNRMI1}}), this argument is not provided and the base measure is a gamma distribution on the common scale parameter.
Traditionally, there is sufficient information in the data to estimate the common scale parameter and inference is not very sensitive to the shape of the base measure.
For the fully nonparametric model, the base measure on the scale parameters can be a gamma, lognormal, half Cauchy, half normal, half Student-t,  uniform or truncated normal distribution.
Additional arguments can be used to tune the shape of the base measure.
\item \texttt{(Alpha, Kappa, Gama)}: Mixing measure parameters identifying a \textbf{Normalised generalised gamma} process, see the L\'evy intensity~\eqref{eq:ngg} with parameters $(\alpha,\kappa,\gamma)$ for more details.
\item The rest of the parameters provide handles to tune the \gls{mcmc} algorithm.
\end{itemize}
Functions to fit a model return an object with \texttt{\hlstd{print}}, \texttt{\hlstd{summary}} and \texttt{\hlstd{plot}} methods, as follows  (the latter plot is represented in Figure~\ref{fig:density-plot}):

\begin{minipage}{\textwidth}
\begin{knitrout}\scriptsize
\definecolor{shadecolor}{rgb}{0.969, 0.969, 0.969}\color{fgcolor}\begin{kframe}
\begin{alltt}
\hlkwd{data}\hlstd{(acidity)}
\hlstd{out} \hlkwb{<-} \hlkwd{MixNRMI1}\hlstd{(acidity)}
\end{alltt}
\begin{verbatim}
## MCMC iteration 500 of 1500 
## MCMC iteration 1000 of 1500 
## MCMC iteration 1500 of 1500 
##  >>> Total processing time (sec.):
##    user  system elapsed 
##  45.840   0.035  45.886
\end{verbatim}
\begin{alltt}
\hlkwd{summary}\hlstd{(out)}
\end{alltt}
\begin{verbatim}
## Density estimation using a Normalized stable process,
## with stability parameter Gamma = 0.4
## 
## A semiparametric normal mixture model was used.
## 
## There were 155 data points.
## 
## The MCMC algorithm was run for 1500 iterations with 10% discarded for burn-in.
## 
## To obtain information on the estimated number of clusters,
##  please use summary(object, number_of_clusters = TRUE).
\end{verbatim}
\end{kframe}
\end{knitrout}
\end{minipage}

\section{Package comparison}
\label{sec:pkg-comparison}

In this section, we discuss in detail the features and functionalities offered in three \R packages addressing \gls{bnp} density estimation, namely: \pkg{BNPdensity}, \pkg{BNPmix} \citep{canale2019bnpmix}, and \pkg{DPpackage} \citep{jara2011DPpackage} ~(\pkg{DPpackage} was removed from the CRAN repository, but former versions are available at \url{https://cran.r-project.org/src/contrib/Archive/DPpackage/}). Since the focus of the present paper is mixture modeling and density estimation, note that other packages relying on \gls{bnp} approaches but tackling other questions such as regression (\pkg{PReMiuM}, \citealp{Liverani2013}, \pkg{Bayesian Regression}, \citealp{karabatsos2015menu}), or meta-analysis (\pkg{bspmma}, \citealp{burr2010bspmma}) are not discussed here. Likewise, non Bayesian approaches are deliberately set aside. Table~\ref{tab:comparison} summarises the comparative study of this section.

\def\mixing{{Mixing measure}}
\def\prior{{Prior characteristics}}
\def\data{{Data}}
\def\vis{{Vis. \& Programming}}

\subsection{Inference algorithm}\label{sec:inf-alg}

Efficient posterior computation for BNP mixture models relies on two types of approaches: marginal or conditional. Marginal methods incorporate analytic integration of infinite dimensional parts of the parameter, which is the case of \pkg{DPpackage} and \pkg{BNPmix}. Instead, \pkg{BNPdensity} relies on a conditional sampler that directly samples trajectories of the processes. More specifically, the Ferguson \& Klass algorithm is employed (see Section~\ref{sec:posterior-sampling}), with the crucial merit of ensuring that largest weights in the series representation are not left out. This is to be compared to the stick-breaking representation where the weights sequence is decreasing only stochastically (that is, in expectation). 

\subsection{Mixing measure}\label{sec:mixing}

As described in Section~\ref{sec:scope}, \gls{bnp} mixture modeling and density estimation require to specify some mixing measure. We start here by comparing the mixing measures available in the three packages. 

\begin{center}
\begin{figure}[hbt]
\centering
  \includegraphics[width = .4\textwidth]{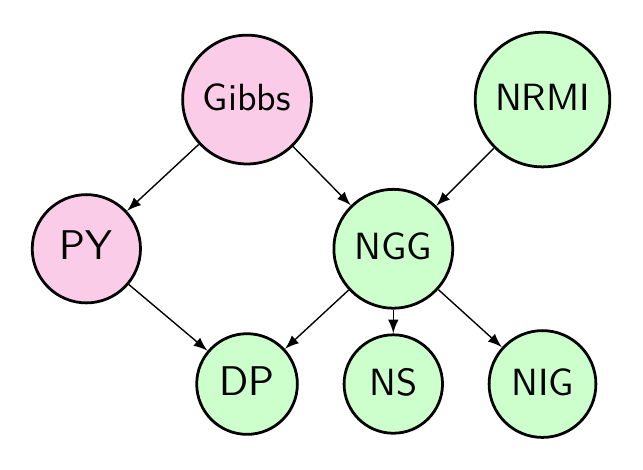}
  \caption{\gls{bnp} priors mentioned in this section. An arrow indicates that the target is a special case or a limit case of its origin. Gibbs: Gibbs-type process. NRMI: normalised random measures with independent increments. NGG: normalised generalised gamma process. PY: Pitman--Yor process. NIG: normalised inverse Gaussian process. NS: normalised stable process. DP: Dirichlet process. In green: covered by \pkg{BNPdensity} package.
  \label{fig:bnp-priors}}
\end{figure}
\end{center}

\pkg{BNPmix} provides a set of functions for density estimation with Dirichlet process and Pitman--Yor mixing measures via marginal algorithms. 
\pkg{DPpackage} is a more general purpose package than both \pkg{BNPdensity} and \pkg{BNPmix}, including functions for regression models, generalised linear mixed models, and generalised additive models, on top of the density model. However, the implementation is primarily tailored to the Dirichlet process mixing measure. 
A natural extension to the Dirichlet and Pitman--Yor processes are Gibbs-type priors \citep{DeBlasi2015}. \glspl{NRMI} are large classes of priors than Gibbs-type priors, and their intersection is identified by the \gls{NGG} process considered in \pkg{BNPdensity}, as established in \cite{lijoi2008investigating}. Being an extremely general class of priors, Gibbs-type processes are beyond reach for a general treatment in a software, however both \pkg{BNPdensity} and \pkg{BNPmix} packages cover its most commonly used sub-classes. 
Pitman--Yor process is not implemented in \pkg{BNPdensity} as it is not an NRMI; yet, a dependence to \pkg{BNPmix} is made in \pkg{BNPdensity}, in such a way that users interested in comparing their results with Pitman--Yor can also use the dedicated functions \code{MixPY1} (semiparametric) and \code{MixPY2} (fully nonparametric) that call \pkg{BNPmix} \code{PYdensity} function. 
The mixing measures covered by the three packages and their mutual relationships are illustrated in Figure~\ref{fig:bnp-priors}.

\subsection{Prior characteristics}\label{sec:prior}

\subsubsection{Non-conjugacy}

Mixture models present the difficulty that the likelihood goes to infinity for infinitely small clusters located exactly on one observed data point.
This may induce numerical problems and instabilities, and such tiny clusters are almost invariably undesirable in practical applications. 
A reasonable solution in the Bayesian framework is to use a prior distribution on scale parameters with little mass on very small values, i.e. a gamma distribution with shape parameter larger than 1 or a truncated distribution.
We might also want to provide a different kind of information on cluster scales: for instance, for a dataset whose variance has been scaled to 1, there is no reason to find clusters with a variance much larger than one.
This would suggest using a prior with an upper bound, or with light tails for large values.
Finally, flexibility in the choice of the kernel $k$ is a clear asset when modelling real data, to choose a reasonable error model.
These three examples suggest that we might need a certain flexibility in the specification of the prior distribution on scale parameters or in the choice of the kernel. 

The inference algorithm used in \pkg{BNPdensity} and presented in Section~\ref{sec:posterior-sampling} does not rely on conjugacy between the base measure and the kernel of the mixture, as do standard algorithms for sampling from a Dirichlet mixture process such as that presented in \cite{escobar1995bayesian}. In contrast, \pkg{DPpackage} and \pkg{BNPmix} are limited to using conjugate couples of base measure and the mixture kernel.

Not being bounded to conjugacy allows us first to use any relevant kernel for the mixture. 
Moreover, even in the case of the normal kernel, this removes the dependence imposed in the conjugate case between the location of the clusters and their variances.
More precisely, this allows a full flexibility on specifying priors based on external knowledge, and proves particularly useful concerning the scale parameters of the kernels.
Indeed, half-Cauchy or half-Gaussian priors for hierarchical variance parameters have recently become popular  \cite{Gelman2006,Chung2015}. 
The illustration on \gls{ssd} (\Cref{sec:case_study}), where the data is scaled, offers such an example where both an upper bound and lower bound on the cluster variances are useful.

\subsubsection{Prior distribution on number of components}

Prior elicitation is a delicate task in Bayesian modeling. \pkg{BNPdensity}  provides some guidelines on how to choose parameters \texttt{(Alpha, Kappa, Gama)} with two functions, one for computing the prior expected number of components, and one for plotting this prior distribution. Comparable functionalities are offered in \pkg{BNPmix} and \pkg{DPpackage}.

The \texttt{(Alpha, Kappa, Gama)} parametrisation allows to easily compare several well known priors. 
We already mentioned that the Dirichlet process can be obtained by setting \texttt{\hlstd{Gama} \hlkwb{=} \hlnum{0}}, the normalised inverse Gaussian process by setting \texttt{Alpha = 1, Gama = 1/2} and the normalised stable process by setting \texttt{Alpha = 1, Kappa = 0}.
The stable process is a convenient model because its parameter $\gamma$ has a simple interpretation: it can be used to tune how informative the prior on the number of components is.
Small values of \texttt{\hlstd{Gama}} bring the process closer to a Dirichlet process, where the prior on the number of components is a relatively peaked distribution around $\alpha\log n$.
In contrast, the larger the value of \texttt{\hlstd{Gama}} is, the flatter the distribution is. 
More guidelines on how to choose the parameters may be found in \cite{lijoi2007controlling}, notably by considering the expected prior number of components.
The expected prior number of components for normalised generalised gamma processes is not trivial to compute due to numerical instabilities, but we provide functions to compute prior distribution on the number of clusters for the normalised stable process and for the Dirichlet process.
These functions require installing the packages \pkg{gmp} and \pkg{Rmpfr} for Multiple Precision Arithmetic, both available on CRAN.

\begin{knitrout}\scriptsize
\definecolor{shadecolor}{rgb}{0.969, 0.969, 0.969}\color{fgcolor}\begin{kframe}
\begin{alltt}
\hlstd{Rmpfr}\hlopt{::}\hlkwd{asNumeric}\hlstd{(}\hlkwd{expected_number_of_components_stable}\hlstd{(}\hlkwc{n} \hlstd{=} \hlnum{100}\hlstd{,} \hlkwc{Gama} \hlstd{=} \hlnum{0.4}\hlstd{))}
\end{alltt}
\begin{verbatim}
## [1] 7.102731
\end{verbatim}
\begin{alltt}
\hlkwd{expected_number_of_components_Dirichlet}\hlstd{(}\hlkwc{n} \hlstd{=} \hlnum{100}\hlstd{,} \hlkwc{Alpha} \hlstd{=} \hlnum{1.}\hlstd{)}
\end{alltt}
\begin{verbatim}
## [1] 5.187378
\end{verbatim}
\end{kframe}
\end{knitrout}

We also provide a way to visualise the prior distribution on the number of components:

\begin{knitrout}\scriptsize
\definecolor{shadecolor}{rgb}{0.969, 0.969, 0.969}\color{fgcolor}\begin{kframe}
\begin{alltt}
\hlkwd{plot_prior_number_of_components}\hlstd{(}\hlnum{100}\hlstd{,} \hlnum{0.4}\hlstd{)}
\end{alltt}
\begin{verbatim}
## Computing the prior probability on the number of clusters for the Dirichlet process
## Computing the prior probability on the number of clusters for the Stable process
\end{verbatim}
\end{kframe}

{\centering \includegraphics[width=\linewidth]{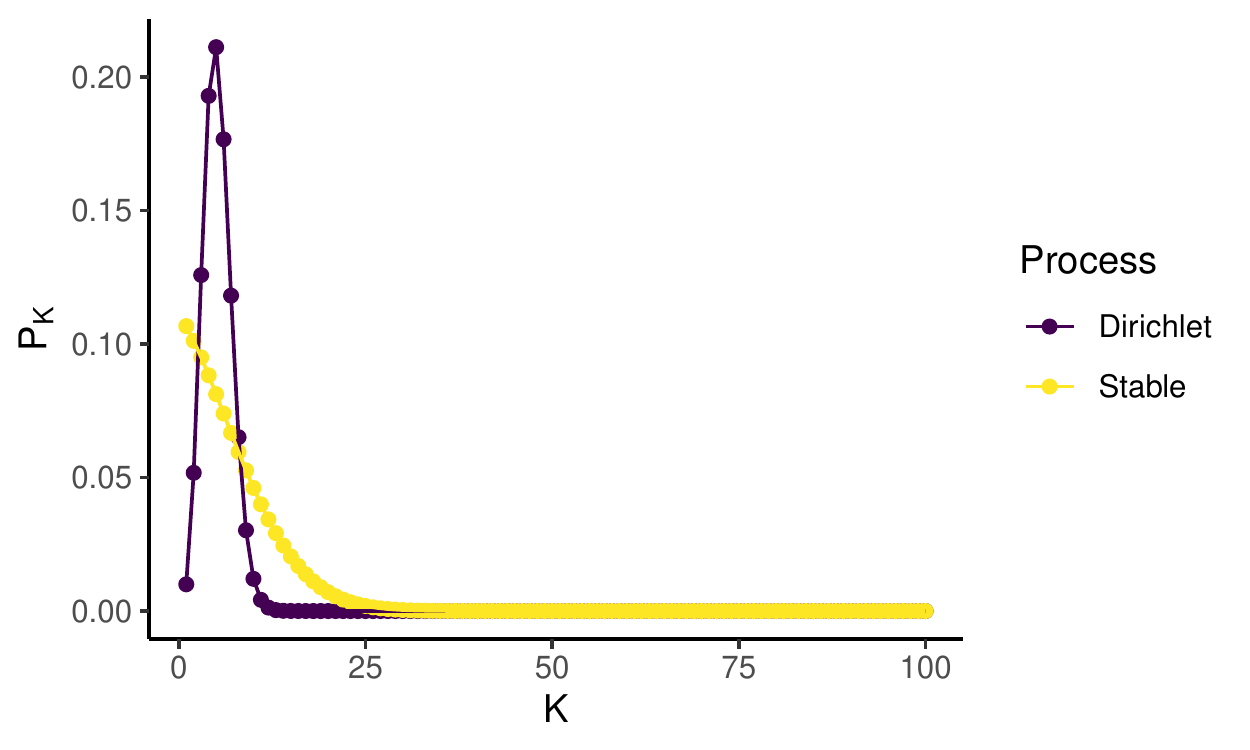} 

}

\end{knitrout}

\captionof{figure}{Prior distribution on the number of clusters with 100 data points, for the stable process with $\gamma = 0.4$ and for the Dirichlet process with $\alpha = 1$.}

\subsection{Censored data}\label{sec:data}

\pkg{BNPdensity} can deal with left, right and interval-censored data by using the functions \code{MixNRMI1cens} and \code{MixNRMI2cens}. The same holds true for \pkg{DPpackage}, while \pkg{BNPmix} does not handle censored data at all.

Censored data usually emerge from imperfections of the measurement process, such as detection limits (high or low) or saturation, low measurement precision, or binning of the data.
Improper treatment of censored data is clearly a source of bias \citep{Helsel2005}: in the case of right-censored data due to a detection limit for high values, for instance, data are not censored at random and discarding them or substituting them deteriorates the dataset.

We deal with censored data by using a version of the likelihood \citep{Helsel2005} adapted to censored data. 
More specifically, denote by $F_k$ the cumulative distribution function of the kernel $k$. 
The heart of the method is then to replace $k(x\mid\theta)$  by $F_k(x\mid\theta)$ for a left-censored observation, by $1-F_k(x\mid\theta)$ for a right-censored observation, and by $F_k(x_r\mid\theta)-F_k(x_l\mid\theta)$ for an interval-censored observation $[x_l,x_r]$.

\subsection{Visualisation and programming}\label{sec:vis}

\subsubsection{Convergence checking and model evaluation}

\pkg{BNPdensity} offers several tools for assessing \gls{mcmc} convergence and performing model checking and comparison.
Notably, we provide a conversion function \texttt{\hlstd{as.mcmc}} to interface the package with the \pkg{coda} package for analysing output and carrying out diagnostics on \gls{mcmc}. We are not aware of such tools for \pkg{BNPmix} or \pkg{DPpackage}.

This is done by running multiple chains starting from different initial conditions, potentially in parallel, and converting them into an \code{mcmc} object that can be processed by \pkg{coda}. A simple solution for running multiple chains does not seem  available for \pkg{BNPmix} and \pkg{DPpackage}.

One conceptual detail for assessing convergence is that, due to the nonparametric nature of the model, the number of parameters which could potentially be monitored to measure auto-correlation of the chains or effective sample size varies. 
The location parameters of the clusters, for instance, vary at each iteration, and even the labels of the clusters vary, which makes it tricky to follow.
However, it is possible to monitor the log-likelihood of the data along the iterations, the value of the latent variable $u$, the number of components and for the semi-parametric model, the value of the common scale parameter.
The following code shows how to compute the potential scale reduction factor \citep{Gelman1992}:

\begin{knitrout}\scriptsize
\definecolor{shadecolor}{rgb}{0.969, 0.969, 0.969}\color{fgcolor}\begin{kframe}
\begin{alltt}
\hlkwd{library}\hlstd{(coda)}
\hlkwd{data}\hlstd{(acidity)}
\hlstd{fit} \hlkwb{=} \hlkwd{multMixNRMI1}\hlstd{(acidity,} \hlkwc{extras} \hlstd{=} \hlnum{TRUE}\hlstd{,} \hlkwc{Nit} \hlstd{=} \hlnum{20000}\hlstd{)}
\hlstd{mcmc_list} \hlkwb{=} \hlkwd{as.mcmc}\hlstd{(fit)}
\hlkwd{gelman.diag}\hlstd{(mcmc_list)}
\end{alltt}
\end{kframe}
\end{knitrout}

\begin{knitrout}\scriptsize
\definecolor{shadecolor}{rgb}{0.969, 0.969, 0.969}\color{fgcolor}\begin{kframe}
\begin{verbatim}
## Potential scale reduction factors:
## 
##                 Point est. Upper C.I.
## ncomp                 1.02       1.06
## Sigma                 1.02       1.07
## Latent_variable       1.02       1.05
## log_likelihood        1.01       1.04
## 
## Multivariate psrf
## 
## 1.03
\end{verbatim}
\end{kframe}
\end{knitrout}

A trace plot for the chains may also be obtained by calling \texttt{\hlkwd{traceplot}\hlstd{(fit)}}; see \Cref{fig:traceplot}.

\begin{center}
 \includegraphics[width=0.5\textwidth]{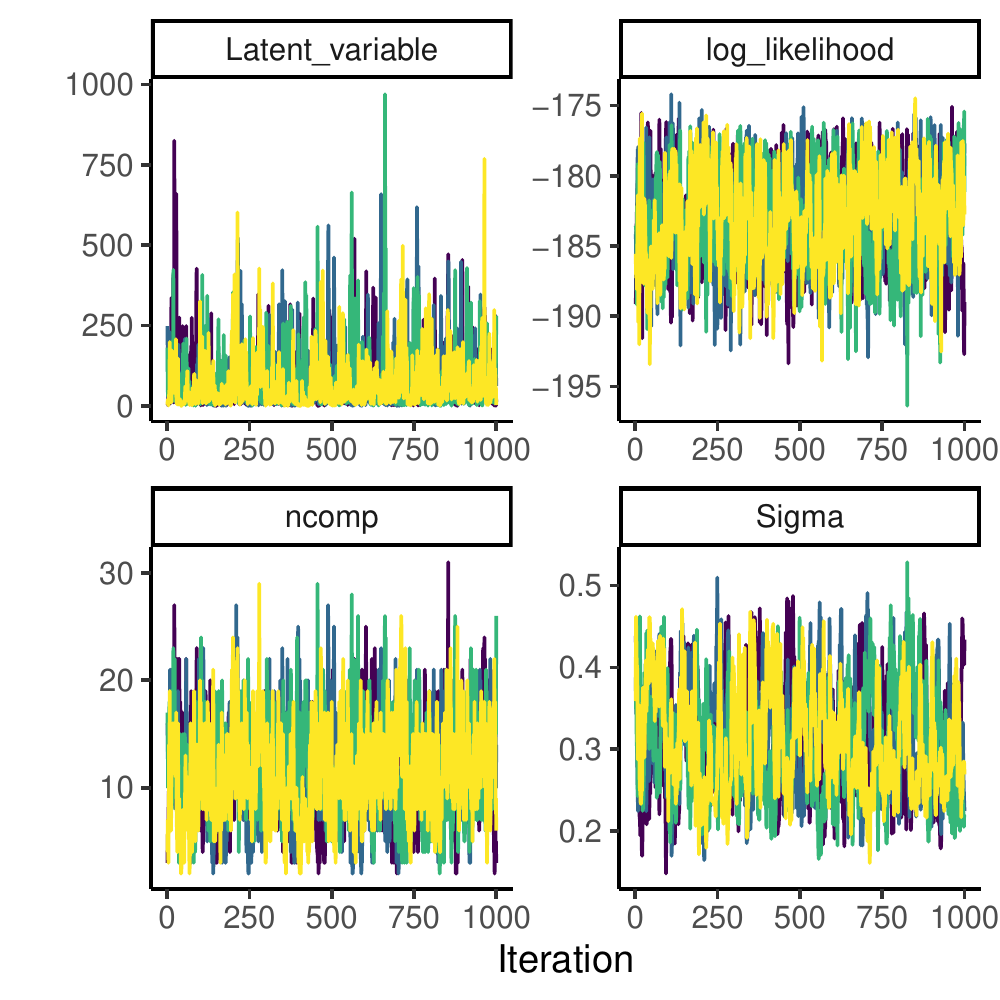}
\captionof{figure}{Trace plot of four chains in the \gls{mcmc} for a semi-parametric model. \label{fig:traceplot}}
\end{center}

\renewcommand{\arraystretch}{1.2}

\begin{table}[hbt]

  \caption{Comparison of \R packages performing \gls{bnp} density estimation: \pkg{BNPdensity}, \pkg{BNPmix}, and \pkg{DPpackage}. (a) See discussion in Section~\ref{sec:mixing}. (b) The \pkg{DPpackage} \code{LDPDdoublyint} function, for \emph{Linear Dependent Poisson Dirichlet Process Mixture Models for the Analysis of Doubly-Interval-Censored Data} could in principle be used for Pitman--Yor process mixture density estimation, although the interface (and the name) suggests it is not intended for this.
  \label{tab:comparison}}

  \begin{tabular}{llccc} 
  \toprule
   & & \pkg{BNPdensity} & \pkg{BNPmix} & \pkg{DPpackage} \\
	\midrule 
	\multirow{2}{*}{\ref{sec:inf-alg} Inference algorithm} 
  & Conditional & \yess & \nos & \nos \\
  & Marginal & \nos & \yess & \yess \\
    \hline 
\multirow{5}{*}{\ref{sec:mixing} \mixing} 
  & Dirichlet process (DP) & \yess & \yess & \yess \\
  & Norm. inverse Gaussian (NIG) & \yess & \nos & \nos \\
  & Norm. stable (NS) & \yess & \nos & \nos \\
  & Norm. gener. gamma (NGG) & \yess & \nos & \nos \\
  & Pitman--Yor (PY) & \nos$^{\text{(a)}}$ & \yess & \nos$^{\text{(b)}}$\\
\hline 
\multirow{2}{*}{\ref{sec:prior} \prior} 
  & Non Gaussian kernels allowed & \yess & \nos & \nos \\
  & Functions for prior elicitation & \yess & \yess & \yess \\
\hline 
\multirow{1}{*}{\ref{sec:data} \data} 
 & All types of censored data & \yess & \nos & \yess \\
\hline 
\multirow{4}{*}{\ref{sec:vis} \vis} 
  & MCMC conv. assessm. & \yess & \nos & \nos \\
  & Graphical model checking & \yess & \nos & \nos \\
  & Clustering vis. tools & \yess & \nos & \nos \\
  & Parallel computing & \yess & \nos & \nos \\
\bottomrule 
  \end{tabular}

\end{table}

We also provide tools for assessing goodness of fit.
Graphical assessment can be performed comparing various representations of the estimated distributions against representations of the empirical distribution (\Cref{GOFplot}). 
Such plots may be obtained from a fitted object using the command  \code{GOFplots(fit, qq\_plot = TRUE)}.
The density plot provides a familiar representation of the Nonparametric distribution, while the \gls{CDF} plot is probably the most classical visualisation of goodness of fit. 
The percentile-percentile plot focuses on the goodness of fit in the center of the distribution, while the quantile-quantile plot focuses on the goodness of fit in the tails of the distribution.
The density, \gls{CDF}, percentile and quantiles used in the plots are the expected posterior quantities, computed from the MCMC sample.
Computation of the theoretical quantiles is a fairly expensive operation because it requires numerically inverting the \gls{CDF}.
We choose not to compute the quantile-quantile plot by default, and when we do, the computation is done on a thinned MCMC chain with an argument provided to control the level of thinning.
\begin{minipage}[c]{\textwidth}
\begin{minipage}[c]{0.45\textwidth}
\includegraphics[width = \textwidth]{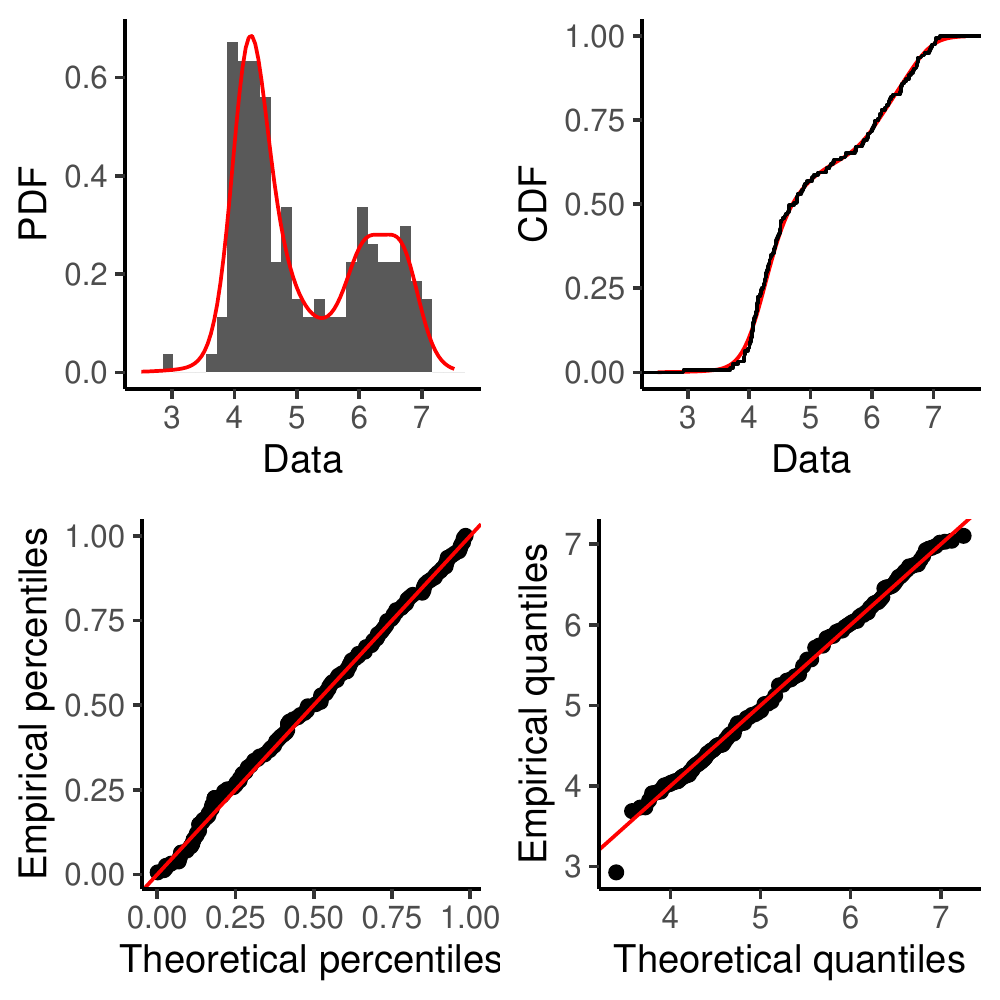}
\end{minipage}
\begin{minipage}[c]{0.45\textwidth}
\includegraphics[width = \textwidth]{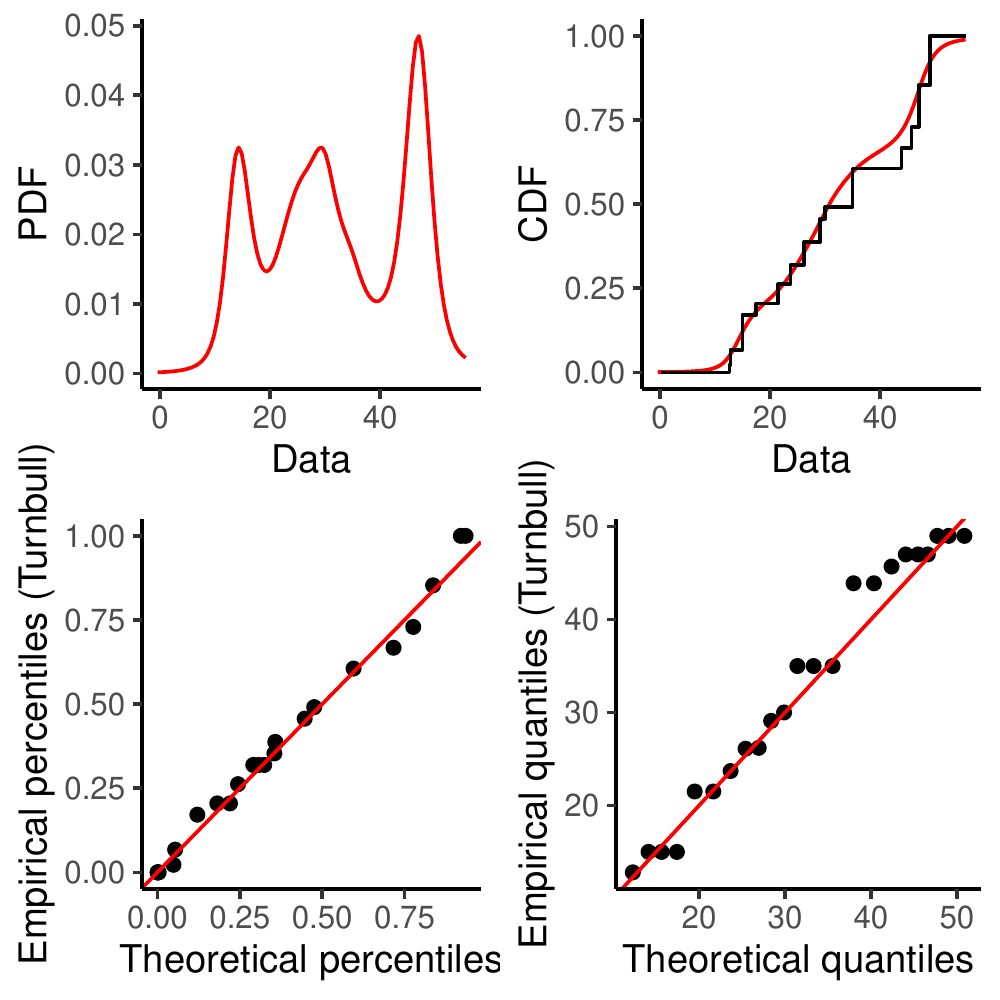}
\end{minipage}
\captionof{figure}{Graphical goodness of fit plots for censored (right) and non censored data (left).
The top row is the mean density estimate with a histogram for the non censored data.
The middle row is the estimated \gls{CDF} with the empirical \gls{CDF} for non censored data, and with the Turnbull estimate of the \gls{CDF} for censored data.
The bottom row are percentile-percentile plots where the empirical percentiles are computed from the empirical \gls{CDF} for the non censored data, and from the Turnbull estimate for the censored data.\label{GOFplot}}\bigskip
\end{minipage}
We also provide tools for model comparison based on expected predictive density.
The \gls{cpo} is the expected predictive density of a data point given the prior and all other data points, so it is the leave-one-out expected predictive density of the model \citep{gelman2013bayesian}, a typical cross-validation criterion.
As such, it is a measure of predictive power with a penalisation for over-fitting.
A Monte Carlo approximation of the \gls{cpo} is easily available and can be used to compare a semi-parametric model to the fully nonparametric model for instance:

\begin{knitrout}\scriptsize
\definecolor{shadecolor}{rgb}{0.969, 0.969, 0.969}\color{fgcolor}\begin{kframe}
\begin{alltt}
\hlkwd{set.seed}\hlstd{(}\hlnum{0}\hlstd{)}
\hlstd{normal_mixture} \hlkwb{<-} \hlkwd{MixNRMI2}\hlstd{(acidity,} \hlkwc{distr.k} \hlstd{=} \hlnum{1}\hlstd{,} \hlkwc{Nit} \hlstd{=} \hlnum{15000}\hlstd{)}
\hlstd{dbl_exponential_mixture} \hlkwb{<-} \hlkwd{MixNRMI1}\hlstd{(acidity,} \hlkwc{distr.k} \hlstd{=} \hlnum{4}\hlstd{,} \hlkwc{Nit} \hlstd{=} \hlnum{15000}\hlstd{)}
\hlkwd{c}\hlstd{(}\hlkwd{median}\hlstd{(normal_mixture}\hlopt{$}\hlstd{cpo),} \hlkwd{median}\hlstd{(dbl_exponential_mixture}\hlopt{$}\hlstd{cpo))}
\end{alltt}
\end{kframe}
\end{knitrout}
\begin{knitrout}\scriptsize
\definecolor{shadecolor}{rgb}{0.969, 0.969, 0.969}\color{fgcolor}\begin{kframe}
\begin{verbatim}
## [1] 0.279 0.271
\end{verbatim}
\end{kframe}
\end{knitrout}

\begin{knitrout}\scriptsize
\definecolor{shadecolor}{rgb}{0.969, 0.969, 0.969}\color{fgcolor}
\begin{tabular}{l|r|r}
\hline
Model & Mean CPO & Median CPO\\
\hline
Nonparametric normal mixture & 0.362 & 0.279\\
\hline
Semi parametric double exponential mixture & 0.357 & 0.271\\
\hline
\end{tabular}

\end{knitrout}

\subsubsection{Clustering visualisation tools}

As described in Section~\ref{sec:clustering}, \pkg{BNPdensity} provides functions for clustering estimation, \code{compute\_optimal\_clustering}, and visual representation, \code{plot\_clustering\_and\_CDF}. 
See also Figure~\ref{fig:clustering_example} and Figure~\ref{fig:clustering_Carbaryl_cens} for illustrations. We are not aware of such clustering tools for \pkg{BNPmix} or \pkg{DPpackage}.

\section{Case study: Species Sensitivity Distribution} \label{sec:case_study}

We present an application of nonparametric density estimation for environmental data.

Assessing the response of a community of species to an environmental stress is of critical importance for ecological risk assessment. 
Methods for this purpose vary in levels of complexity and realism. 
\gls{ssd} represents an intermediate tier, more refined than rudimentary assessment factors \citep{Posthuma2010} but practical enough for routine use by environmental managers and regulators in most developed countries (Australia, Canada, China, EU, South Africa, USA,\ldots).
The \gls{ssd} approach is intended to provide, for a given contaminant, a description of the tolerance of all species possibly exposed using information collected on a sample of those species. 
This information consists of a single species-specific value, which marks a limit over which the species suffers adverse effects.  
This value is very often censored \citep{KonKamKing2014}, because measuring it is both costly and difficult (bioassay experiments).
The tolerance of all species possibly exposed is described by  a distribution, fitted on the sample of species \citep{Aldenberg2000d}.
The quantity of interest for ecological risk assessment is the \gls{hc5} , which
corresponds to the \nth{5} percentile of the \gls{ssd} distribution.
The lack of justification for the choice of any given parametric distribution has sparked several research directions. 
Some authors \citep{Xu2015, He2014, Jagoe1997, VanStraalen2002, Xing2014, Zhao2016} have sought to find the best parametric distribution by model comparison using goodness-of-fit measures. 
The general understanding is that no single distribution seems to provide a superior fit and that the answer is dataset dependent \citep{Forbes2002}. 
Therefore, the log-normal distribution has become the customary choice, notably because it readily provides confidence intervals on the \gls{hc5}, and because model comparison and goodness of fit tests have relatively low power on small datasets, precluding the emergence of a definite answer to the question. 

The availability of a package such as \pkg{BNPdensity} allows to move beyond this customary assumption very easily.
\glspl{NRMI} offer a flexible nonparametric mixture model, which can accommodate distributions very different from a normal distribution.
\cite{barrios2013modeling} and \cite{KonKamKing2017} show that \glspl{NRMI} have better performance than Dirichlet process mixtures, kernel density estimates (the recent approach proposed by \cite{Wang2015}) or simple one-component normal models.
Moreover, there are good reasons to believe that the distribution of species sensibility should at least allow for multimodality. 
Indeed, many stressors target specifically certain species groups, such as insecticides for insects, while they are developed with the aim of leaving other species group unaffected.
Therefore, it is expected that there should at the very least be a group of sensitive species and a group of less sensitive species.
This is why \cite{Zajdlik2009} propose to model the species sensitivity distribution as a finite mixture, with raises customary issues of model choice.
Using a \gls{bnp} approach via \pkg{BNPdensity} allows generalising this approach while circumventing the theoretical and technical difficulties of estimating the right number of components in a mixture.

It is also important to use a method which may be applied to small datasets. 
This is another motivation for using a \gls{bnp} approach, where model complexity adapts to the number of data points, and will tend to suggest simple or even univariate mixtures when few data points are present.
On the contrary, many classical nonparametric approaches to modelling species sensitivity distribution \citep{Wang2015,Verdonck2001} only work well on large datasets.

To model species sensitivity distribution, we carefully select the parameters in the package \pkg{BNPdensity}.
Given that concentrations vary on a wide range, it is common practice to work
on log-transformed concentrations.
We choose a fully nonparametric model using the normalised stable process \citep{kingman1975random} as mixing random measure (hence setting \code{Alpha} = 1 and \code{Beta} = 0).
We favor this process over the more classical Dirichlet process because it allows specifying less informative prior on the number of components, which makes it more robust to model misspecification \citep{barrios2013modeling}.
With this process, the amount of information from the prior is controlled by the stability parameter $\gamma$, which we set to 0.4 (\code{Gama} = 0.4).
This choice reflects a compromise between model flexibility ($\gamma \to 1$) and computational effort ($\gamma$ small, see also \cref{sec:package}).
As we wish the location parameter of the clusters $\boldsymbol{\mu}$ to be estimated freely, we use the default weakly informative prior of a normal base measure $f_{0}^1(\boldsymbol{\mu}|\boldsymbol{\varphi}) = \mathcal{N}(\boldsymbol{\mu}|\varphi_1, \varphi_2)$ with hyperpriors on $\boldsymbol{\varphi}$ given by $f(\boldsymbol{\varphi}) = \mathcal{N}(\varphi_1|\psi_1, \psi_2)\text{ga}(\varphi_2|\psi_3, \psi_4)$ (see also \cite{barrios2013modeling} for more details).

For the prior on the scale of the clusters, we want to use two pieces of information: first, since the data has been scaled, scale parameters are likely to be smaller than 1, the extreme case being a mixture with a single component. 
Second, we want to avoid the possibility of extremely small clusters centred on a data point, because they are not very interesting from an interpretation point of view, and because they cause numerical problems (the likelihood diverges when a cluster scale goes to 0).
Therefore, we choose a uniform distribution between 0.1 and 1.5 for the prior on the cluster scales.

In keeping with the traditional assumption of normality of the species sensitivity distribution, we choose to use a normal kernel for the mixture ($\code{distr.k} = 1$).

We now compare three approaches to modelling \glsfirst{ssd}: the most standard and recommended approach of \cite{Wagner1991, Aldenberg2000d}, which is a simple normal model, the most recent proposal by \citep{Wang2015} which is a normal kernel density estimate and the \gls{bnp} normal mixture made available with \pkg{BNPdensity} that we presented above. 
 As already stated, a quantity of interest is the \nth{5} percentile of the distribution.
 We choose as an
estimator the median of the posterior distribution of the \nth{5} percentile, while
the 95\% credible bands are formed by the 2.5\% and 97.5\% quantiles of the
posterior distribution of the \nth{5} percentile. 
The \nth{5} percentile of the \gls{kde} is
obtained by numerical inversion of the cumulative distribution function, and
the confidence intervals using the nonparametric bootstrap. 
The \nth{5} percentile
of the normal \gls{ssd} and its confidence intervals are obtained following the
classical method of \cite{Aldenberg2000d}.

We use data from an ecotoxicity research database as pre-processed in \cite{Hickey2012}.
We extract data for the insecticide Carbaryl.
The dataset contains 57 species, of which approximately 40\% have censored data.
We obtain a non censored version of this dataset by excluding right or left censored data, and replacing interval censored data by the midpoint of the interval.  
\cite{helsel2006fabricating, Dowse2013,KonKamKing2014} have shown that transforming censored data risks inducing bias, hence the ability of \pkg{BNPdensity} to accommodate censoring is particularly valuable for \gls{ssd}.
There does not appear to be any easily available approach to use \gls{kde} methods on all types of censored data.
\Cref{fig:appli_atrazine} shows a comparison of three approaches to \gls{ssd}.
The left hand side of \Cref{fig:appli_atrazine} shows that the \gls{bnp} model is more flexible than both the \gls{kde} and normal model, while the right hand side shows that it is no less robust, according to a leave-one-out cross validation criterion.
The middle panel shows that although the \gls{bnp} model is more flexible and takes into account uncertainty on the number of clusters, the estimation of the \nth{5} percentile is not much more uncertain than with the other methods. Significantly larger uncertainty would have jeopardised the real world applicability of the \gls{bnp}-\gls{ssd}.

\begin{center}
 \includegraphics[width=0.32\textwidth]{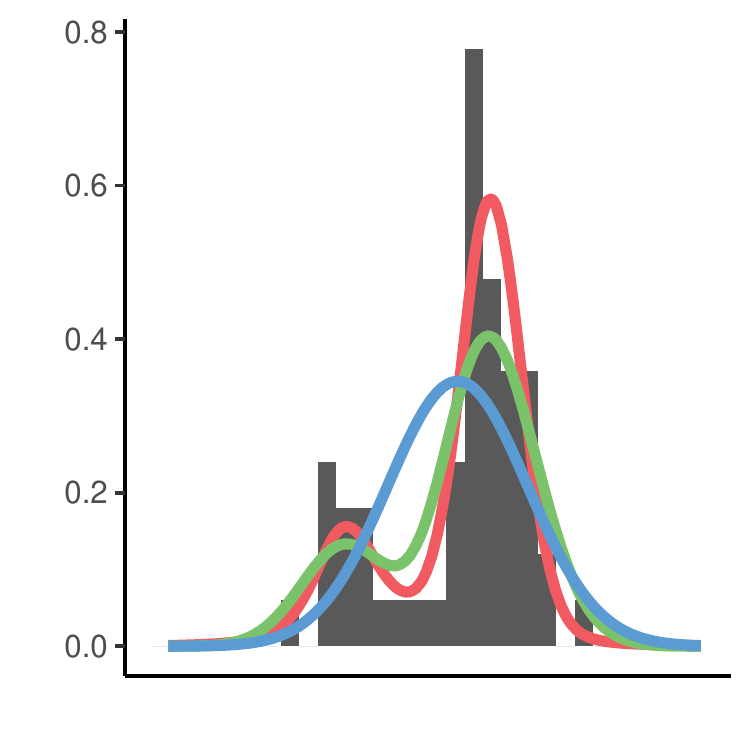}
 \includegraphics[width=0.32\textwidth]{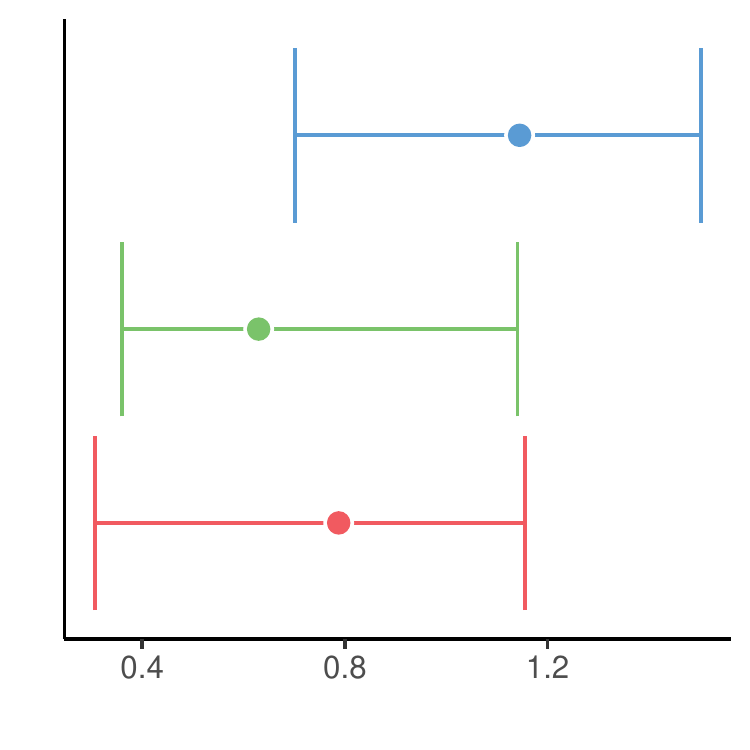}
 \includegraphics[trim={0cm 1cm 0cm 0cm},clip,width=.32\textwidth]{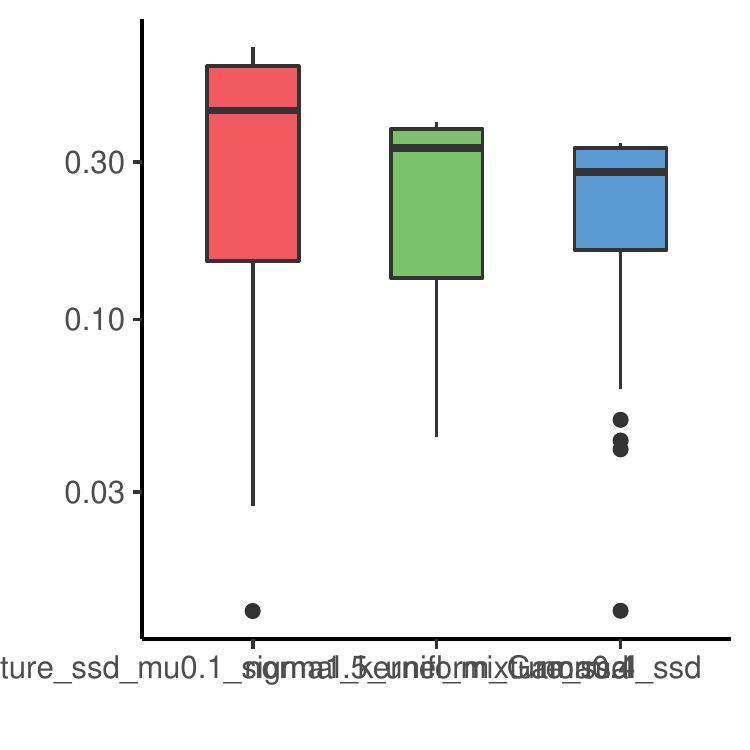}
 \includegraphics[width=0.32\textwidth]{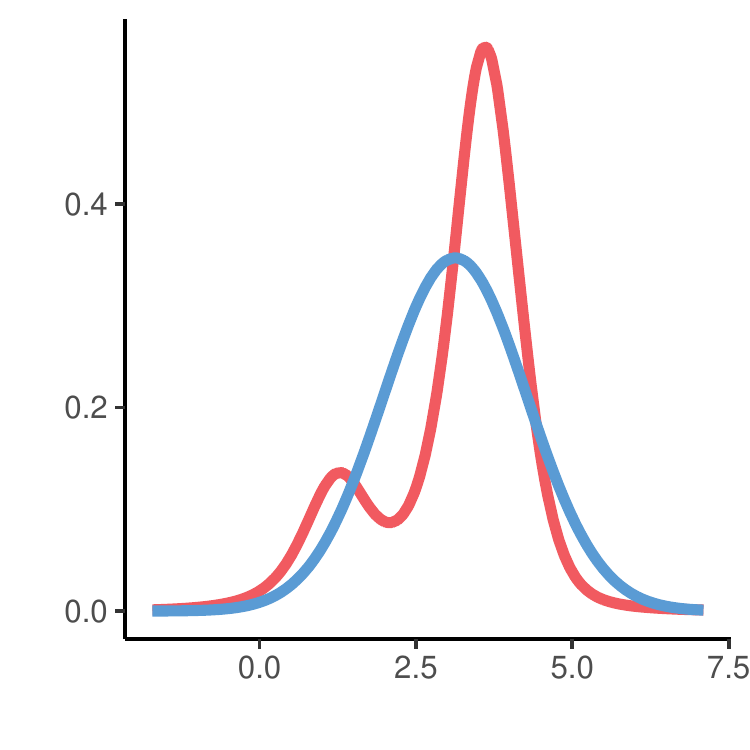}
 \includegraphics[width=0.32\textwidth]{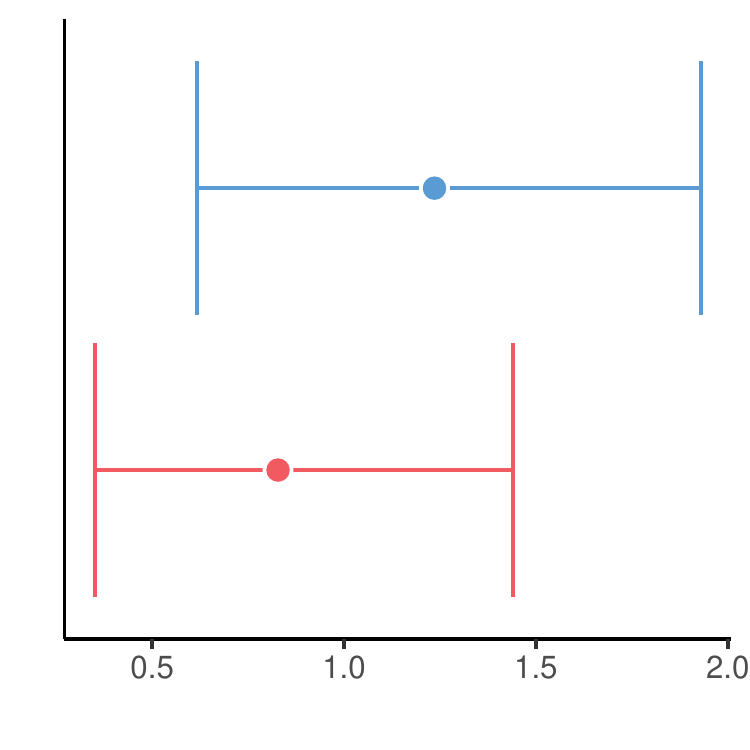}
 \includegraphics[width=0.32\textwidth]{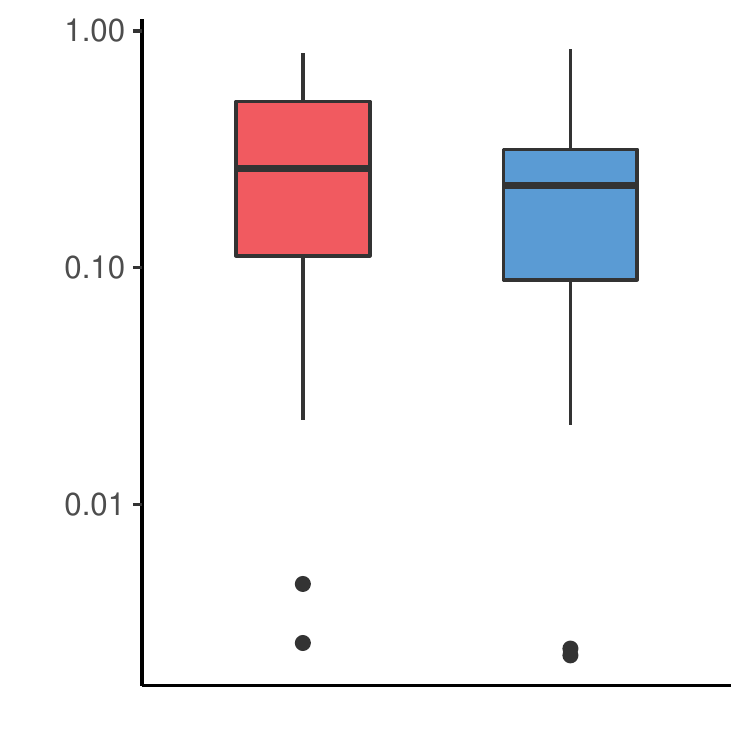}
\captionof{figure}{Top panel: non censored data. Bottom panel: censored data. The normal model is represented in blue, the \gls{kde} in green and the BNP in red. 
Left: density plot and histogram for the Carbaryl data using several \gls{ssd} methods. The histogram is not available for censored data.
Center: \nth{5} percentile estimate (not available for \gls{kde} with censored data). Right: boxplot of the \gls{cpo} (for BNP) and \gls{loo} (for normal and \gls{kde}, not available for \gls{kde} with censored data), one value for each data point. \label{fig:appli_atrazine}}
\end{center}

An added value of the \gls{bnp}-\gls{ssd} is that on top of being more flexible than the classic normal \gls{ssd} and more robust than the nonparametric approach of \cite{Wang2015}, as a mixture model it naturally induces a clustering of the data which may contain some biologically interesting information.
We implemented functions to estimate the optimal clustering from the \gls{mcmc} sample and visualise it, potentially including a label on each point to reflect available meta data for interpretation.
In the context of \gls{ssd}, it is interesting to know what drives species sensitivity: it might be taxonomy, in the sense that taxonomically close species will tend to respond in the same way and belong to the same cluster, but other drivers have been suggested such as habitat, feeding behaviour or respiration, which may not coincide with taxonomy.
\Cref{fig:clustering_Carbaryl_cens} shows the clustering induced in the case of the insecticide Carbaryl.
In this case, there is a large cluster mostly composed of fish and molluscs, and a cluster mostly composed of insects and crustaceans, showing that the clustering structure is consistent with a finer taxonomic structure. This suggests that for Carbaryl, taxonomy may very well be the main driver for sensitivity.

\begin{center}
	\includegraphics[width = .6\textwidth]{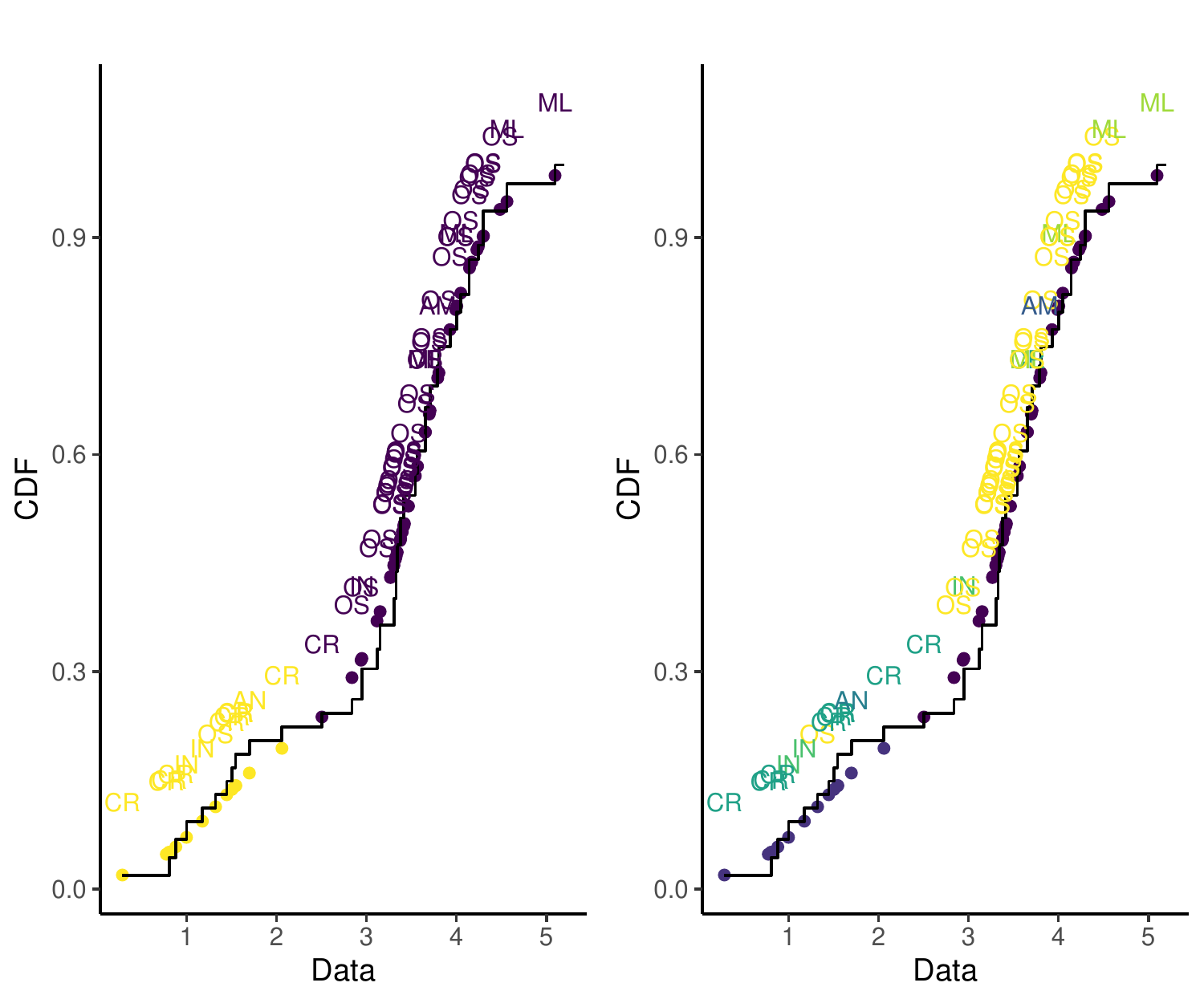}
\captionof{figure}{Graphical representation of the clustering induced by the mixture model for the Carbaryl data. 
The solid line represents the Turnbull estimate of the \gls{CDF}, the points loosely represent the data.
Interval censored data are represented at the middle of the interval, left and right censored data are not represented.
A label describing the taxonomic group of each species is written above each point, AM: Amphibians, AN: Annelids (worms), CR: Crustaceans, IN: Insects, ML: Molluscs, OS: Osteichthyes (fish). 
On the left panel, the points and the labels are coloured according to the estimated cluster index.
On the right panel, the labels are coloured according to the taxonomic group and the points are not coloured.
\label{fig:clustering_Carbaryl_cens}}
\end{center}

\section*{Computational details}

The results in this paper were obtained using
\R~4.1.1 with the
\BNPdensity package version 2020.3.4. \R itself
and all packages used are available from the Comprehensive
\R Archive Network (CRAN) at
\url{https://CRAN.R-project.org/}.

\section*{Acknowledgments}
We would like to thank Matti Vihola for useful advice on adaptive MCMC.
J. Arbel is  partially supported by MIAI@Grenoble Alpes (ANR-19-P3IA-0003).
A. Lijoi and I. Pr\"unster are partially supported by MIUR, PRIN Project
2015SNS29B.

\newpage
\bibliographystyle{apalike}
\bibliography{jabref_lib}

\begin{thebibliography}{}

\bibitem[Aldenberg and Jaworska, 2000]{Aldenberg2000d}
Aldenberg, T. and Jaworska, J.~S. (2000).
\newblock {Uncertainty of the hazardous concentration and fraction affected for
  normal species sensitivity distributions.}
\newblock {\em Ecotoxicology and Environmental Safety}, 46(1):1--18.

\bibitem[Arbel and Pr{\"{u}}nster, 2017]{arbel2017moment}
Arbel, J. and Pr{\"{u}}nster, I. (2017).
\newblock {A moment-matching Ferguson {\&} Klass algorithm}.
\newblock {\em Statistics and Computing}, 27(1):3--17.

\bibitem[Barrios et~al., 2013]{barrios2013modeling}
Barrios, E., Lijoi, A., Nieto-Barajas, L.~E., and Pr{\"u}nster, I. (2013).
\newblock Modeling with normalized random measure mixture models.
\newblock {\em Statistical Science}, 28(3):313--334.

\bibitem[Binder, 1978]{binder1978bayesian}
Binder, D.~A. (1978).
\newblock {Bayesian cluster analysis}.
\newblock {\em Biometrika}, 65(1):31--38.

\bibitem[Brix, 1999]{brix1999generalized}
Brix, A. (1999).
\newblock Generalized gamma measures and shot-noise {Cox} processes.
\newblock {\em Advances in Applied Probability}, 31:929--953.

\bibitem[Burr, 2012]{burr2010bspmma}
Burr, D. (2012).
\newblock {bspmma: An R package for Bayesian semi-parametric models for
  metaanalysis}.
\newblock {\em Journal of Statistical Software}, 50(4):1--23.

\bibitem[Bush and MacEachern, 1996]{bush}
Bush, C.~A. and MacEachern, S.~N. (1996).
\newblock {A semiparametric Bayesian model for randomised block designs}.
\newblock {\em Biometrika}, 83(2):275--285.

\bibitem[Canale et~al., 2019]{canale2019bnpmix}
Canale, A., Corradin, R., and Nipoti, B. (2019).
\newblock {BNPmix: an R package for Bayesian nonparametric modelling via
  Pitman--Yor mixtures}.
\newblock {\em Journal of Statistical Software}, page to appear.

\bibitem[Chung et~al., 2015]{Chung2015}
Chung, Y., Gelman, A.~G., Rabe-Hesketh, S., Liu, J., and Dorie, V. (2015).
\newblock {Weakly Informative Prior for Point Estimation of Covariance Matrices
  in Hierarchical Models}.
\newblock {\em Journal of Educational and Behavioral Statistics},
  40(2):136--157.

\bibitem[Dahl, 2006]{dahl2006model}
Dahl, D.~B. (2006).
\newblock {Model-based clustering for expression data via a Dirichlet process
  mixture model}.
\newblock {\em Bayesian inference for gene expression and proteomics},
  4:201--218.

\bibitem[{De Blasi} et~al., 2015]{DeBlasi2015}
{De Blasi}, P., Favaro, S., Lijoi, A., Mena, R.~H., Pr{\"{u}}nster, I., and
  Ruggiero, M. (2015).
\newblock {Are Gibbs-type priors the most natural generalization of the
  Dirichlet process?}
\newblock {\em IEEE Transactions on Pattern Analysis and Machine Intelligence},
  37(2):212--229.

\bibitem[Denwood, 2016]{denwood2016runjags}
Denwood, M.~J. (2016).
\newblock {runjags: An R package providing interface utilities, model
  templates, parallel computing methods and additional distributions for MCMC
  models in JAGS}.
\newblock {\em Journal of Statistical Software}, 71(9):1--25.

\bibitem[Dowse et~al., 2013]{Dowse2013}
Dowse, R., Tang, D., Palmer, C.~G., and Kefford, B.~J. (2013).
\newblock {Risk assessment using the species sensitivity distribution method:
  Data quality versus data quantity}.
\newblock {\em Environmental Toxicology and Chemistry}, 32(6):1360--1369.

\bibitem[Escobar and West, 1995]{escobar1995bayesian}
Escobar, M.~D. and West, M. (1995).
\newblock {Bayesian Density Estimation and Inference Using Mixtures}.
\newblock {\em Journal of the American Statistical Association},
  90(430):577--588.

\bibitem[Ferguson and Klass, 1972]{ferguson1972representation}
Ferguson, T.~S. and Klass, M.~J. (1972).
\newblock {A representation of independent increment processes without Gaussian
  components}.
\newblock {\em Ann. Math. Stat.}, 43(5):1634--1643.

\bibitem[Forbes and Calow, 2002]{Forbes2002}
Forbes, V.~E. and Calow, P. (2002).
\newblock {Species Sensitivity Distributions Revisited: A Critical Appraisal}.
\newblock {\em Human and Ecological Risk Assessment}, 8(3):473--492.

\bibitem[Fr\"uhwirth-Schnatter et~al., 2018]{handbook2018}
Fr\"uhwirth-Schnatter, S., Celeux, G., and Robert, C.~P. (2018).
\newblock {\em {Handbook of Mixture Analysis}}.
\newblock Chapman \& Hall/CRC.

\bibitem[Gelman, 2006]{Gelman2006}
Gelman, A.~G. (2006).
\newblock {Prior distributions for variance parameters in hierarchical models
  (Comment on Article by Browne and Draper)}.
\newblock {\em Bayesian Analysis}, 1(3):515--534.

\bibitem[Gelman et~al., 2014]{gelman2013bayesian}
Gelman, A.~G., Carlin, J.~B., Stern, H.~S., and Rubin, D.~B. (2014).
\newblock {\em {Bayesian Data Analysis}}.
\newblock CRC press, Boca Raton, FL, third edition.

\bibitem[Gelman and Rubin, 1992]{Gelman1992}
Gelman, A.~G. and Rubin, D.~B. (1992).
\newblock {Inference from Iterative Simulation Using Multiple Sequences}.
\newblock {\em Statistical Science}, 7(4):457--511.

\bibitem[Gilks et~al., 1993]{gilks1994language}
Gilks, W.~R., Thomas, A., and Spiegelhalter, D.~J. (1993).
\newblock {A Language and program for complex bayesian modelling}.
\newblock {\em Journal of the Royal Statistical Society. Series D (The
  Statistician)}, 43(1, Special):169--177.

\bibitem[He et~al., 2014]{He2014}
He, W., Qin, N., Kong, X., Liu, W., Wu, W., He, Q., Yang, C., Jiang, Y., Wang,
  Q., Yang, B., and Xu, F. (2014).
\newblock {Ecological risk assessment and priority setting for typical toxic
  pollutants in the water from Beijing-Tianjin-Bohai area using Bayesian
  matbugs calculator (BMC)}.
\newblock {\em Ecological Indicators}, 45:209--218.

\bibitem[Helsel, 2005]{Helsel2005}
Helsel, D.~R. (2005).
\newblock {\em {Nondetects and data analysis. Statistics for censored
  environmental data}}.
\newblock Wiley-Interscience.

\bibitem[Helsel, 2006]{helsel2006fabricating}
Helsel, D.~R. (2006).
\newblock Fabricating data: how substituting values for nondetects can ruin
  results, and what can be done about it.
\newblock {\em Chemosphere}, 65(11):2434--2439.

\bibitem[Hickey et~al., 2012]{Hickey2012}
Hickey, G.~L., Craig, P.~S., Luttik, R., and de~Zwart, D. (2012).
\newblock {On the quantification of intertest variability in ecotoxicity data
  with application to species sensitivity distributions}.
\newblock {\em Environmental Toxicology and Chemistry}, 31(8):1903--1910.

\bibitem[Jagoe and Newman, 1997]{Jagoe1997}
Jagoe, R.~H. and Newman, M.~C. (1997).
\newblock {Bootstrap estimation of community NOEC values}.
\newblock {\em Ecotoxicology}, 6(5):293--306.

\bibitem[James et~al., 2009]{james2009posterior}
James, L.~F., Lijoi, A., and Pr{\"{u}}nster, I. (2009).
\newblock {Posterior analysis for normalized random measures with independent
  increments}.
\newblock {\em Scandinavian Journal of Statistics}, 36(1):76--97.

\bibitem[Jara, 2007]{jara2007applied}
Jara, A. (2007).
\newblock Applied {B}ayesian non- and semi-parametric inference using
  {DPpackage}.
\newblock {\em R News}, 7(3):17--26.

\bibitem[Jara et~al., 2011]{jara2011DPpackage}
Jara, A., Hanson, T.~E., Quintana, F.~A., M{\"{u}}ller, P., and Rosner, G.~L.
  (2011).
\newblock {DPpackage: Bayesian non-and semi-parametric modelling in R}.
\newblock {\em Journal of statistical software}, 40(5):1.

\bibitem[Karabatsos, 2017]{karabatsos2015menu}
Karabatsos, G. (2017).
\newblock A menu-driven software package of bayesian nonparametric (and
  parametric) mixed models for regression analysis and density estimation.
\newblock {\em Behavior Research Methods}, 49:335--362.

\bibitem[Kingman, 1975]{kingman1975random}
Kingman, J. (1975).
\newblock Random discrete distributions.
\newblock {\em Journal of the Royal Statistical Society. Series B},
  37(1):1--15.

\bibitem[{Kon Kam King} et~al., 2017]{KonKamKing2017}
{Kon Kam King}, G., Arbel, J., and Pr{\"{u}}nster, I. (2017).
\newblock {A Bayesian Nonparametric Approach to Ecological Risk Assessment}.
\newblock In Argiento, R., Lanzarone, E., {Antoniano Villalobos}, I., and
  Mattei, A., editors, {\em Bayesian Statistics in Action: BAYSM 2016,
  Florence, Italy, June 19-21}, pages 151--159. Springer International
  Publishing, Cham.

\bibitem[{Kon Kam King} et~al., 2014]{KonKamKing2014}
{Kon Kam King}, G., Veber, P., Charles, S., and Delignette-Muller, M.~L.
  (2014).
\newblock {MOSAIC{\_}SSD: A new web tool for species sensitivity distribution
  to include censored data by maximum likelihood}.
\newblock {\em Environmental Toxicology and Chemistry}, 33(9):2133--2139.

\bibitem[Lau and Green, 2007]{lau2007bayesian}
Lau, J.~W. and Green, P.~J. (2007).
\newblock Bayesian model-based clustering procedures.
\newblock {\em Journal of Computational and Graphical Statistics},
  16(3):526--558.

\bibitem[Lijoi et~al., 2005]{lijoi05}
Lijoi, A., Mena, R.~H., and Pr{\"u}nster, I. (2005).
\newblock {Hierarchical mixture modeling with normalized inverse-Gaussian
  priors}.
\newblock {\em Journal of the American Statistical Association},
  100(472):1278--1291.

\bibitem[Lijoi et~al., 2007a]{lijoi07}
Lijoi, A., Mena, R.~H., and Pr{\"u}nster, I. (2007a).
\newblock {Bayesian nonparametric estimation of the probability of discovering
  new species}.
\newblock {\em Biometrika}, 94(4):769--786.

\bibitem[Lijoi et~al., 2007b]{lijoi2007controlling}
Lijoi, A., Mena, R.~H., and Pr{\"u}nster, I. (2007b).
\newblock {Controlling the reinforcement in Bayesian non-parametric mixture
  models}.
\newblock {\em J. Roy. Stat. Soc. B Met.}, 69(4):715--740.

\bibitem[Lijoi et~al., 2008]{lijoi2008investigating}
Lijoi, A., Pr{\"{u}}nster, I., and Walker, S.~G. (2008).
\newblock {Investigating nonparametric priors with Gibbs structure}.
\newblock {\em Statistica Sinica}, 18(4):1653--1668.

\bibitem[Liverani et~al., 2015]{Liverani2013}
Liverani, S., Hastie, D.~I., Azizi, L., Papathomas, M., and Richardson, S.
  (2015).
\newblock {PReMiuM}: An {R} package for profile regression mixture models using
  {D}irichlet processes.
\newblock {\em Journal of Statistical Software}, 64(7):1--30.

\bibitem[Lo, 1984]{lo1984class}
Lo, A. (1984).
\newblock {On a class of Bayesian nonparametric estimates: I. Density
  estimates}.
\newblock {\em The Annals of Statistics}, 12(1):351--357.

\bibitem[MacEachern and M{\"{u}}ller, 1998]{MacEachern1998}
MacEachern, S.~N. and M{\"{u}}ller, P. (1998).
\newblock {Estimating Mixture of Dirichlet Process Models}.
\newblock {\em Journal of Computational and Graphical Statistics},
  7(2):223--238.

\bibitem[Meila, 2007]{meila2007comparing}
Meila, M. (2007).
\newblock Comparing clusterings---an information based distance.
\newblock {\em Journal of Multivariate Analysis}, 98(5):873--895.

\bibitem[Neal, 2000]{neal2000markov}
Neal, R.~M. (2000).
\newblock {Markov chain sampling methods for Dirichlet process mixture models}.
\newblock {\em Journal of computational and graphical statistics},
  9(2):249--265.

\bibitem[Papaspiliopoulos and Roberts, 2008]{papaspiliopoulos2008retrospective}
Papaspiliopoulos, O. and Roberts, G. (2008).
\newblock {Retrospective Markov chain Monte Carlo methods for Dirichlet process
  hierarchical models}.
\newblock {\em Biometrika}, 95(1):169.

\bibitem[Plummer, 2003]{plummer2003jags}
Plummer, M. (2003).
\newblock {JAGS: a program for analysis of Bayesian graphical models using
  Gibbs sampling}.
\newblock In {\em Proceedings of the 3rd International Workshop on Distributed
  Statistical Computing (DSC 2003), March 20-22, Vienna, Austria. ISSN
  1609-395X.}, volume 124, page 125.

\bibitem[Plummer, 2019]{rjags2019plummer}
Plummer, M. (2019).
\newblock {\em {rjags: Bayesian Graphical Models using MCMC}}.
\newblock R package version 4-9.

\bibitem[Posthuma et~al., 2002]{Posthuma2010}
Posthuma, L., {Suter II}, G.~W., and Trass, P.~T. (2002).
\newblock {\em {Species sensitivity distributions in ecotoxicology}}.
\newblock CRC press.

\bibitem[Rastelli and Friel, 2018]{Rastelli2018}
Rastelli, R. and Friel, N. (2018).
\newblock {Optimal Bayesian estimators for latent variable cluster models}.
\newblock {\em Statistics and Computing}, 28(6):1169--1186.

\bibitem[RCoreTeam, 2019]{rcoreteam}
RCoreTeam (2019).
\newblock {R: A Language and Environment for Statistical Computing}.

\bibitem[Regazzini et~al., 2003]{rlp2003}
Regazzini, E., Lijoi, A., and Pr{\"{u}}nster, I. (2003).
\newblock {Distributional results for means of normalized random measures with
  independent increments}.
\newblock {\em Annals of Statistics}, 31(2):560--585.

\bibitem[Roberts and Rosenthal, 2009]{roberts2009examples}
Roberts, G.~O. and Rosenthal, J.~S. (2009).
\newblock Examples of adaptive mcmc.
\newblock {\em Journal of Computational and Graphical Statistics},
  18(2):349--367.

\bibitem[Sato, 1999]{sato1999levy}
Sato, K.-I. (1999).
\newblock {\em L\'evy Processes and Infinitely Divisible Distributions},
  volume~68 of {\em Cambridge {{Studies}} in {{Advanced Mathematics}}}.
\newblock {Cambridge University Press}.

\bibitem[Scrucca et~al., 2016]{Scrucca2016}
Scrucca, L., Fop, M., Murphy, T.~B., and Raftery, A.~E. (2016).
\newblock {mclust} 5: clustering, classification and density estimation using
  {G}aussian finite mixture models.
\newblock {\em The {R} Journal}, 8(1):205--233.

\bibitem[{Stan Development Team}, 2018]{rstan-software}
{Stan Development Team} (2018).
\newblock {RStan}: the {R} interface to {Stan}.
\newblock R package version 2.18.2.

\bibitem[{Stan Development Team} and {Stan Developement Team},
  2019]{stan-software:2015}
{Stan Development Team} and {Stan Developement Team} (2019).
\newblock {Stan: A C++ Library for Probability and Sampling, Version 2.19}.

\bibitem[Sturtz et~al., 2005]{sturtz2005r2winbugs}
Sturtz, S., Ligges, U., and Gelman, A.~E. (2005).
\newblock {R2WinBUGS: a package for running WinBUGS from R}.
\newblock {\em Journal of Statistical Software}, 12(3):1--16.

\bibitem[Thomas et~al., 2006]{thomas2006making}
Thomas, A., O'Hara, B., Ligges, U., and Sturtz, S. (2006).
\newblock {Making BUGS open}.
\newblock {\em R News}, 6(1):12--17.

\bibitem[Todeschini et~al., 2014]{Biips}
Todeschini, A., Caron, F., and Fuentes, M. (2014).
\newblock {Rbiips: Bayesian inference with interacting particle systems}.
\newblock {\em arXiv}.

\bibitem[{Van Straalen}, 2002]{VanStraalen2002}
{Van Straalen}, N.~M. (2002).
\newblock {Threshold models for species sensitivity distributions applied to
  aquatic risk assessment for zinc}.
\newblock {\em Environmental Toxicology and Pharmacology}, 11(3-4):167--172.

\bibitem[Verdonck et~al., 2001]{Verdonck2001}
Verdonck, F. A.~M., Jaworska, J., Thas, O., and Vanrolleghem, P.~A. (2001).
\newblock {Determining environmental standards using bootstrapping, Bayesian
  and maximum likelihood techniques: A comparative study}.
\newblock {\em Analytica Chimica Acta}, 446(1-2):429--438.

\bibitem[Wade and Ghahramani, 2018]{wade2018bayesian}
Wade, S. and Ghahramani, Z. (2018).
\newblock {Bayesian cluster analysis: Point estimation and credible balls (with
  discussion)}.
\newblock {\em Bayesian Analysis}, 13(2):559--626.

\bibitem[Wagner and Lokke, 1991]{Wagner1991}
Wagner, C. and Lokke, H. (1991).
\newblock {Estimation of ecotoxicological protection levels from NOEC toxicity
  data}.
\newblock {\em Water Research}, 25(10):1237--1242.

\bibitem[Wang et~al., 2015]{Wang2015}
Wang, Y., Wu, F., Giesy, J.~P., Feng, C., Liu, Y., Qin, N., and Zhao, Y.
  (2015).
\newblock {Non-parametric kernel density estimation of species sensitivity
  distributions in developing water quality criteria of metals}.
\newblock {\em Environmental Science and Pollution Research},
  22(18):13980--13989.

\bibitem[Xing et~al., 2014]{Xing2014}
Xing, L., Liu, H., Zhang, X., Hecker, M., Giesy, J.~P., and Yu, H. (2014).
\newblock {A comparison of statistical methods for deriving freshwater quality
  criteria for the protection of aquatic organisms}.
\newblock {\em Environmental Science and Pollution Research}, 21(1):159--167.

\bibitem[Xu et~al., 2015]{Xu2015}
Xu, F.-L., Li, Y.-L., Wang, Y., He, W., Kong, X.-Z., Qin, N., Liu, W.-X., Wu,
  W.-J., and Jorgensen, S.~E. (2015).
\newblock {Key issues for the development and application of the species
  sensitivity distribution (SSD) model for ecological risk assessment}.
\newblock {\em Ecological Indicators}, 54:227--237.

\bibitem[Zajdlik et~al., 2009]{Zajdlik2009}
Zajdlik, B.~A., Dixon, D.~G., and Stephenson, G. (2009).
\newblock {Estimating Water Quality Guidelines for Environmental Contaminants
  Using Multimodal Species Sensitivity Distributions: A Case Study with
  Atrazine}.
\newblock {\em Human and Ecological Risk Assessment}, 15(3):554--564.

\bibitem[Zhao and Chen, 2016]{Zhao2016}
Zhao, J. and Chen, B. (2016).
\newblock {Species sensitivity distribution for chlorpyrifos to aquatic
  organisms: Model choice and sample size.}
\newblock {\em Ecotoxicology and Environmental Safety}, 125:161--9.

\end{thebibliography}

\end{document}